\documentclass[12pt]{article}
\pdfoutput=1
\usepackage{fullpage}
\usepackage{amsmath}
\usepackage{amssymb}
\usepackage{bbm}
\usepackage{color}
\usepackage{ulem}

\newcommand{\Comment}[1]{{}}
\definecolor{MyDarkBlue}{rgb}{0.15,0.15,0.45}
\usepackage[linktocpage=true]{hyperref}
\hypersetup{
colorlinks=true,
citecolor=MyDarkBlue,
linkcolor=MyDarkBlue,
urlcolor=MyDarkBlue,
pdfauthor={},
pdftitle={},
pdfsubject={hep-th}
}

\DeclareFontFamily{OT1}{rsfs10}{}
\DeclareFontShape{OT1}{rsfs10}{m}{n}{ <-> rsfs10 }{}
\DeclareMathAlphabet{\mathscript}{OT1}{rsfs10}{m}{n}

\def\gsim{ \lower .75ex \hbox{$\sim$} \llap{\raise .27ex \hbox{$>$}} }
\def\lsim{ \lower .75ex \hbox{$\sim$} \llap{\raise .27ex \hbox{$<$}} }
\def\be{\begin{equation}}
\def\ee{\end{equation}}
\def\bea{\begin{eqnarray}}
\def\eea{\end{eqnarray}}

\newcommand{\ns}{\normalsize}

\def\({\left(}
\def\){\right)}

\newcommand{\pt}{\partial}

\usepackage{latexsym,amsmath,amssymb,epsfig}

\usepackage{graphicx}
\usepackage{graphicx,subfigure}
\usepackage{epstopdf}
\usepackage[body={17.5cm, 21cm},right=2cm]{geometry}
\usepackage{amssymb}
\usepackage{amsmath}
\usepackage{psfrag}
\usepackage{epsfig}
\usepackage{cancel}
 \allowdisplaybreaks[4]

\usepackage[all]{xy}

\begin{document}

\begin{titlepage}


\title{
  \hfill{\ns }  \\
   {\LARGE The Pseudo-Conformal Universe: Scale Invariance from Spontaneous Breaking of Conformal Symmetry}
\\
}
\author{
   Kurt Hinterbichler and Justin Khoury
     \\
{\ns Center for Particle Cosmology, Department of Physics \& Astronomy} \\[-0.4cm]
{\ns  University of Pennsylvania, Philadelphia, PA 19104}\\
[0.3cm]}

\date{}

\maketitle

\begin{abstract}
We present a novel theory of the very early universe which addresses the traditional horizon and flatness problems of big bang cosmology and predicts a scale invariant spectrum of perturbations.  Unlike inflation, this scenario requires no exponential accelerated expansion of space-time.  Instead, the early universe is described by a conformal field theory minimally coupled to gravity.  The conformal fields develop a time-dependent expectation value which breaks the flat space $so(4,2)$ conformal symmetry down to $so(4,1)$, the symmetries of de Sitter, giving perturbations a scale invariant spectrum. The solution is an attractor, at least in the case of a single time-dependent field. Meanwhile, the metric background remains approximately flat but slowly contracts, which makes the universe increasingly flat, homogeneous and isotropic, akin to the smoothing mechanism of ekpyrotic cosmology. 
Our scenario is very general, requiring only a conformal field theory capable of developing the appropriate time-dependent expectation values, and encompasses existing incarnations of this idea, specifically the 
$U(1)$ model of Rubakov and the Galileon Genesis scenario.  Its essential features depend only on the symmetry breaking pattern and not on the details of the underlying lagrangian. It makes generic observational predictions that make it potentially distinguishable from standard inflation, in particular significant non-gaussianities and the absence of primordial gravitational waves.
\end{abstract}

\end{titlepage}
\setcounter{page}{2}

\section{Introduction}

Big Bang cosmology gives a well tested model of the history of our universe from the time of nucleosynthesis until now, but leaves several theoretical problems unaddressed.  These include the horizon problem (why is the universe so uniform across areas of the sky that were causally disconnected at early times?), the flatness problem (why is the universe seen to be spatially flat, even though any initial curvature should tend to dominate with time?), and the problem of the origin of fluctuations (why do we observe a nearly scale invariant spectrum of perturbations in the cosmic microwave background?).  Inflation is the most widely studied theory which addresses these problems within a single framework~\cite{inf}. A common feature of all inflationary models is that early in its history the universe underwent a period of exponential growth, before transitioning into the standard power-law expansion of big bang cosmology.

Though observational evidence for primordial density perturbations with nearly scale invariant and gaussian statistics is compatible with the simplest inflationary models, a critical question is whether inflation is unique in making these predictions. This has motivated cosmologists to devise alternative scenarios, such string gas cosmology~\cite{Brandenberger:1988aj}$-$\cite{Brandenberger:2008nx}, pre-big bang scenarios~\cite{Gasperini:1992em}$-$\cite{Gasperini:2007vw}, ekpyrotic/cyclic theories~\cite{Khoury:2001wf}$-$\cite{Joyce:2011ta} (see~\cite{Lehners:2008vx,Lehners:2010fy} for reviews) and the matter-bounce scenario~\cite{Wands:1998yp}$-$\cite{Cai:2011zx}. The question of uniqueness is particularly timely with the coming of Planck data, which will test the assumptions of scale invariance and gaussianity to unprecedented accuracy. 

In this paper, we develop a novel framework for early universe cosmology, the {\it pseudo-conformal universe}, which addresses the traditional problems of standard big bang cosmology, does not involve an exponentially expanding space-time, and is distinguishable from inflation.  The scenario postulates that the early universe is governed by an approximate conformal field theory (CFT), living in approximately flat space. The algebra of conformal transformations of 4d Minkowksi space is $so(4,2)$.  In the early universe, before the big bang, one or more of the conformal fields develops a specific time-dependent expectation value, which breaks the conformal symmetry down to an $so(4,1)$ subalgebra. This unbroken algebra is isomorphic to the symmetry algebra of de Sitter space, and so certain fields coupled to the CFT will acquire scale invariant spectra as though they were effectively living on an inflating de Sitter space. The smooth scalar field quickly comes to dominate over any spatial curvature components or anisotropies, thereby addressing the horizon and flatness problems.

The scenario is very general.  All that is required is that the early universe be described by a CFT, a state with high symmetry.  In addition, there must be at least one field of conformal weight zero, as these are the ones which acquire a scale invariant spectrum.  The salient features of the scenario depend only on the symmetry breaking pattern $so(4,2)\rightarrow so(4,1)$, and not on the precise details of the underlying microscopic theory.  Furthermore, gravity is not important at early times, unlike inflation, so the scenario can be realized at strong coupling through the AdS/CFT correspondence~\cite{AdSCFT}, which provides an ample set of examples of 4d CFT's.

The general scenario we describe makes use of a CFT with any number of scalar fields $\phi_I$, indexed by $I = 1,\ldots, N$, possessing arbitrary conformal weights $d_I$.  The lagrangian should be such that the fields develop a time-dependent expectation value  $\phi_I (t) \sim {1/ (-t)^{d_I}}$, with $-\infty < t < 0$. As we will see, this accomplishes the desired symmetry breaking pattern $so(4,2)\rightarrow so(4,1)$. The field content and breaking pattern will be enough to fix the form of the quadratic action for fluctuations. In turn, this quadratic action will allow us to show that the background is a dynamical attractor in the case of a single time-dependent field, and calculate the spectrum of gaussian fluctuations, all without knowledge of the explicit lagrangian.  The conformal weight 0 fields will acquire scale invariant spectra, under very general conditions.

The general framework developed in this paper greatly extends existing incarnations of this idea. The $U(1)$ model of Rubakov~\cite{rub1}$-$\cite{rub7} is one example. The model of Galilean Genesis~\cite{galgen} is another. These specific examples underscore that the mechanism is both generic and can be realized in myriad ways. In generalizing, we will see that it is possible to find other models with certain advantages.  

For illustrative purposes, our fiducial example is that of a single weight 1 canonical scalar rolling down an upside down $\phi^4$ potential, first considered by~\cite{rub1}:
\be
{\cal L}_\phi = -\frac{1}{2}(\partial\phi)^2 + \frac{\lambda}{4}\phi^4,
\label{phi4intro}
\ee
with $\lambda>0$.  This theory is (classically) conformally invariant.  As the scalar rolls down the potential, it develops a time dependent expectation value $\phi(t)\sim 1/ t$, which breaks the conformal symmetry in the desired manner.  Coupling to gravity, the background evolution is that of a slowly {\it contracting} universe with a very stiff equation of state $w\gg 1$.  Coupling in a weight 0 field produces the desired scale invariant spectrum.  

It is worth stressing that there are important differences between the model studied by~\cite{rub1} and the cosmological scenario proposed here: $i$) the model
of~\cite{rub1} assumes that the scalar field couples conformally to gravity, whereas we will couple $\phi$ minimally to gravity, thereby weakly
breaking conformal invariance; $ii)$ the complex scalar is a spectator field in~\cite{rub1}, whereas $\phi$ will drive the cosmological background in our case,
resulting in a phase of slow contraction that makes the universe increasingly smooth.

The quartic example~(\ref{phi4intro}) also differs from the Galileon Genesis model in the following ways: $i)$ the background evolution derives from 
a renormalizable scalar field theory, rather than the non-renormalizable galileons; $ii)$ perturbations can be made to propagate at the speed of light, as opposed to the superluminal propagation in Galilean Genesis, thus making the scenario more amenable to a UV completion in quantum field theory or perturbative string theory~\cite{nimaUVIR}; $iii)$ the mechanism does not have to rely on a violation the Null Energy Condition (NEC) during the generation of density perturbations. (Though, as we will see, the scenario takes place in a slowly contracting universe, so the NEC must eventually be violated to obtain a cosmological bounce to an expanding phase.)  

The scenario incorporates many elements of ekpyrotic cosmology, such as the idea of a slowly contracting phase prior to the big bang to smooth out the universe~\cite{Gratton:2003pe,Creminelli:2004jg,Erickson:2003zm,Garfinkle:2008ei} and the use of multiple interacting scalar fields to generate scale invariant perturbations~\cite{Notari:2002yc,Finelli:2002we,Lehners:2007ac,Buchbinder:2007ad,Creminelli:2007aq}. But the scenario also capitalizes on key novel ingredients, such as the role of symmetries and the connection between conformal invariance and scale invariance of the primordial spectrum~\cite{mottola1,mottola2}.

The pseudo-conformal scenario offers significant advantages over existing alternatives to inflation, such as ekpyrotic or cyclic cosmologies~\cite{Steinhardt:2001st}.
Existing alternatives usually rely on specific forms for the scalar potential and highly special couplings between scalar fields to generate a scale invariant spectrum. 
In contrast, our scenario is very general and can be realized by a variety of microphysical lagrangians, as already exemplified by the known
examples mentioned above. The conformal galileon and Rubakov's $U(1)$ model rely on drastically different lagrangians, yet both yield the
desired symmetry breaking pattern. 

Moreover, scale invariance follows from symmetry considerations and does not require finely tuned couplings between scalars. 
{\it Any} conformal weight 0 field will acquire a scale invariant spectrum, under very general assumptions. Finally, existing non-inflationary
mechanisms are generally unstable~\cite{dust,Khoury:2008wj} or can only generate a limited range of modes without hitting strong coupling issues~\cite{Khoury:2010gw,Baumann:2011dt}.
(One exception is the superluminal scenario of~\cite{csm}$-$\cite{Piao:2008ip}, but in this case one must be willing to accept superluminality as physically plausible.)
In known non-inflationary multi-field examples, in particular, scale invariance hinges on a tachyonic instability along the entropy direction in field space~\cite{Koyama:2007mg,Tolley:2007nq},
which inevitably makes the background solution a repellor\footnote{To be precise, the existence or non-existence of such an instability has not been established in~\cite{Creminelli:2007aq}. However, the entropy perturbation grows as $\delta s \sim 1/(-t)$ in this model, just as in~\cite{Buchbinder:2007ad}, which suggests that the background is also tachyonically unstable in this case.}. In contrast, in our case perturbations in the conformal weight 0 field grow to a constant value, as in inflation, hence the background remains an attractor even in the presence of entropic fields.

In Section~\ref{symmbreak}, we introduce the scenario in the general case, a CFT containing any number of scalar fields of various conformal weights where a specific time dependent expectation value is responsible for the symmetry breaking. For illustrative purposes, in Section~\ref{Explicitsection} we explore in detail the explicit negative $\phi^4$ realization. 
In Section~\ref{gravon}, we include gravity to derive the resulting cosmological background, showing how the traditional cosmological problems are addressed, and study the generation of perturbations. Section~\ref{obscons} presents the observational consequences and signatures of pseudo-conformal cosmology. Quantum corrections are the focus of Section~\ref{loopestimates},
where it is shown that the negative $\phi^4$ theory minimally coupled to gravity is radiatively stable. We then consider some other possible realizations of our scenario in Section~\ref{confgal}, comparing to models which have appeared before in the literature as special cases. We consider further extensions of our scenario in Section~\ref{DBIsec}, focusing in particular
on the DBI relativistic extension of the $\phi^4$ example. We summarize our results and discuss future research avenues in Section~\ref{conclusions}.  

\section{Generic Properties of Breaking Conformal Symmetry to de Sitter}
\label{symmbreak}

The conformal algebra of four dimensional Minkowski space is $so(4,2)$.  Our mechanism for generating scale invariant perturbations relies on the spontaneous breaking
of global conformal invariance.  We are interested in the case where the symmetry breaking pattern is
\be
so(4,2)\rightarrow so(4,1)\,.
\ee
The unbroken algebra $so(4,1)$ is isomorphic to the isometry algebra of four dimensional de Sitter space.

We will see in this Section that the essential features of the mechanism, such as stability of the background solution and scale invariance of perturbations,
are completely determined by this symmetry breaking pattern, irrespective of the details of the underlying dynamics. 
For this purpose, we can ignore gravity and work in flat, Minkowski space. We will see later that this is a consistent
approximation at sufficiently early times.

We imagine that the early universe is described by a general conformal field theory involving some collection of conformal scalars $\phi_I$, indexed by $I = 1,\ldots, N$.  Each scalar may have a different conformal weight $d_I$.  The 15 (linear) generators of the conformal transformations act on $\phi_I$ as
\bea 
\nonumber
\delta_{P_\mu}\phi_I &=&-\partial_\mu\phi_I \,, \qquad  \,\,\,\,\,\,\,\,\,\,\,\,\,\,\, \delta_{J^{\mu\nu}}\phi_I  = (x^\mu\partial^\nu-x^\nu\partial^\mu)\phi_I \,,\\
\delta_D\phi_I &=&- (d_I+ x^\mu\partial_\mu) \phi_I \,, \qquad  \delta_{K_\mu}\phi_I  = \(-2x_\mu d_I -2x_\mu x^\nu\partial_\nu +x^2\partial_\mu\)\phi_I \,.
\label{delphiconfmany}
\eea
The $P_\mu$ and $J_{\mu\nu}$ generate space-time translations and rotations, respectively, $D$ generates dilatations, and the
$K^\mu$ generate special conformal transformations. These satisfy the commutation relations
\bea 
\nonumber
\left[\delta_D,\delta_{P_\mu}\right]&=&-\delta_{P_\mu}\,, \qquad\,\,\,\,\,\,\,\,\,\,\,\,\,\,\,\,\,\,\,\, \left [\delta_D,\delta_{K_\mu}\right] = \delta_{K_\mu}\,, \qquad \left[\delta_{K_\mu},\delta_{P_\nu}\right] = 2\(\delta_{J_{\mu\nu}}-\eta_{\mu\nu}\delta_D\)\,,  \\
\nonumber
\left[\delta_{J_{\mu\nu}},\delta_{K_\lambda}\right] &=& \eta_{\lambda\mu}\delta_{K_{\nu}}-\eta_{\lambda\nu}\delta_{K_\mu}\,,\,\,\,\,\,\,\,\, \left[\delta_{J_{\mu\nu}}, \delta_{P_\sigma} \right] = \eta_{\mu\sigma} \delta_{P_\nu}-\eta_{\nu\sigma} \delta_{P_\mu}\,,\\
\left[\delta _{J_{\mu\nu}}, \delta _{J_{\sigma\rho}} \right]&=&\eta _{\mu\sigma} \delta _{J_{\nu\rho}}- \eta _{\nu\sigma} \delta _{J_{\mu\rho}}+\eta _{\nu\rho} \delta _{J_{\mu\sigma}}- \eta _{\mu\rho} \delta _{J_{\nu\sigma}}\,,
\label{nonzerocommutators}
 \eea
with all other commutators being zero. By defining $\delta_{J^{-2,-1}}=\delta_D$, $\delta_{J^{-2,\mu}}=(\delta_{P^\mu}-\delta_{K^\mu})/2$ and $\delta_{J^{-1,\mu}}= (\delta_{P^\mu}+\delta_{K^\mu})/2$, we can assemble all the conformal generators into an anti-symmetric matrix $\delta _{J^{AB}}$, with $A,B$ taking the six values $(-2,-1,\mu)$.  The commutation relations~(\ref{nonzerocommutators}) then take the form
\be 
\left[\delta _{J_{AB}}, \delta _{J_{CD}} \right]=\eta _{AC} \delta _{J_{BD}}- \eta _{BC} \delta _{J_{AD}}+\eta _{BD} \delta _{J_{AC}}- \eta _{AD} \delta _{J_{BC}}\,,
\label{SO(4,2)alg}
\ee
where $\eta_{AB}={\rm diag}(-1,1,\eta_{\mu\nu})$ is a 6d Minkowski metric with two time directions. These are the commutation relations of $so(4,2)$.

The lagrangian should be invariant under all these symmetries.  Constructing the lagrangian out of Poincar\'e invariant terms which all have dimension 4 (so that there are no dimensionful couplings) will guarantee invariance under the Poincar\'e generators as well as under the dilatation generator $\delta_D$.  Invariance under the special conformal generators $\delta_{K_\mu}$ is a more restrictive condition, and is generally not satisfied unless specific linear combinations of dilatation invariant terms are used.  In any case, the details of the lagrangian are not too important, and the features of our scenario follow largely from the symmetry breaking pattern.

Now, suppose the $\phi_I$'s take a time-dependent background value
\be
\bar \phi_I (t) = {c_I\over (-t)^{d_I}}\,,
\label{phibackgenmany}
\ee
where the $c_I$'s are constant coefficients.  Since we envision a pre-big bang scenario, we have chosen time to run over negative values $-\infty < t < 0$.  We claim that if at least one field has $d_I\not=0$ and a non-vanishing background profile, this profile will induce the symmetry breaking pattern~(\ref{symmbreak}).
Indeed, the subalgebra of 10 generators that annihilate $\bar\phi_I$ are 
\be
\delta_{P_i}\,,\ \ \ \delta_D\,,\ \ \ \delta_{J_{ij}}\,,\ \ \ \delta_{K_i}\,,\ \ \ i=1,2,3\,.
\ee
That is, the background~(\ref{phibackgenmany}) preserves spatial translations and rotations, the spatial components of the special conformal transformations, and dilatations.
These 10 unbroken generators can be assembled into an anti-symmetric matrix $\delta _{J^{ab}}$, with $a,b$ taking the five values $(-2,-1,i)$, by defining
$\delta_{J^{-2,-1}}=\delta_D$, $\delta_{J^{-2,i}}= (\delta_{P^i}-\delta_{K^i})/2$ and $\delta_{J^{-1,i}}= (\delta_{P^i}+\delta_{K^i})/2$. 
The commutation relations then take the form
\be 
\left[\delta _{J_{ab}}, \delta _{J_{cd}} \right]=\eta _{ac} \delta _{J_{bd}}- \eta _{bc} \delta _{J_{ad}}+\eta _{bd} \delta _{J_{ac}}- \eta _{ad} \delta _{J_{bc}}\,,
\label{SO(4,1)alg}
\ee
where $\eta_{ab}={\rm diag}(-1,1,\delta_{ij})$ is a 5d Minkowski metric with one time direction.  These are the commutation relations of $so(4,1)$. 
Meanwhile, the remaining 5 generators do not annihilate $\bar\phi_I$, and no non-trivial linear combination of them does. The 5 broken generators act on the background solution as
\be
\delta_{P_0}\bar\phi_I = \frac{d_I\bar\phi_I}{t} \,,\qquad  \delta_{J^{0i}}\bar\phi_I = -\frac{d_Ix^i}{t} \bar\phi_I\,,\qquad \delta_{K_0}\bar\phi_I = -\frac{d_Ix_\mu x^\mu}{t}\bar\phi\,.
\ee
Hence, as claimed, we have the symmetry breaking pattern $so(4,2) \rightarrow so(4,1)$. (In the case where all $d_I = 0$, or all the background profiles vanish for $d_I\not=0$, the full conformal group is preserved.)

\subsection{Effective Action for the Goldstone fluctuations}
\label{quadgold}

We will now see how far this symmetry breaking pattern can take us, without knowing the underlying lagrangian.  We will study the quadratic action for the perturbations (Goldstone fields\footnote{It is well-known that the usual counting of Goldstone modes does not work for spontaneously broken space-time
symmetries~\cite{Low:2001bw}. Though our background breaks 5 symmetries, we only have one Goldstone field.}) around the background (\ref{phibackgenmany}), $\varphi _I=\phi_I -\bar\phi_I$.  We will see that it is fixed by the symmetries up to a few constants.  The action will linearly realize the unbroken symmetries, and non-linearly realize the broken ones.  In this Section we derive the most general, 2-derivative action for $\varphi_I$. We will then use this general action to study the stability of the background (Section~\ref{dynattr}), as well as to derive the spectrum of perturbations generated from an adiabatic vacuum (Sections~\ref{pertspec} and~\ref{conf0}).

The unbroken $so(4,1)$ subalgebra acts linearly on the perturbations $\varphi_I$,
\bea
\nonumber
\delta_{P_i}\varphi_I &=&-\partial_i\varphi_I \,, \qquad\,\,\,\,\,\,\,\,\,\,\,\,\, \delta_{J^{ij}}\varphi_I  =\(x^i\partial^j-x^j\partial^i\)\varphi_I \,, \\
\delta_D\varphi_I &=&-\(d_I+ x^\mu\partial_\mu\)\varphi_I \,,\qquad \delta_{K_i}\varphi_I =\(-2x_id_I -2x_i x^\nu\partial_\nu +x^2\partial_i\)\varphi_I \,, \label{unbrokengenmany}
\eea
whereas the 5 broken generators act non-linearly,
\bea
\nonumber
\delta_{P_0}\varphi_I & =& {d_I\over t}\bar\phi_I-\dot\varphi_I \,,\qquad \delta_{J^{0i}}\varphi_I =-{d_I x^i\over t}\bar\phi_I+\(t\partial_i+x^i\partial_t\)\varphi_I \,, \\
\delta_{K_0}\varphi_I &=&-{d_I x^2 \over t}\bar\phi_I+\(2t d_I+ 2t x^\nu\partial_\nu +x^2\partial_t\)\varphi_I \,.
\label{brokengenmany}
\eea
By imposing these symmetry transformations, we can deduce the form of the quadratic action for the $\varphi_I$'s.  (As mentioned above, the case where all $d_I = 0$ or where all fields with $d_I\not=0$ have a vanishing background profile is exceptional, since all 15 symmetries are linearly realized in this case.)

Consider the original lagrangian ${\cal L}$, and replace the original fields $\phi_I$ with $\bar\phi_I+\gamma \varphi_I$, where $\gamma$ is a formal parameter to count powers of the perturbations.   Since $\bar\phi_I$ is a solution to the equations of motion of ${\cal L}$, expanding the lagrangian for the perturbations in a formal series in $\gamma$ gives an expansion that starts at order $\gamma^2$:
\be 
{\cal L}=\gamma^2\left({\cal L}_{(2)}+\gamma {\cal L}_{(3)}+\gamma^2{\cal L}_{(4)}+\ldots\right) \ .
\ee
The lagrangian ${\cal L}$ possesses the symmetries (\ref{unbrokengenmany}) and (\ref{brokengenmany}), each of which may also be expanded in powers of $\gamma$,
\be 
\delta\varphi_I=\delta_{(0)}\varphi_I+\gamma \delta_{(1)}\varphi_I\ .
\ee
The statement that $\delta\varphi_I$ is a symmetry of ${\cal L}$ is
\be 
\label{invarianceeq} {\delta^{\rm EL}{\cal L}\over \delta \varphi_I}\delta\varphi_I \simeq 0 \ ,
\ee
where $\delta^{\rm EL}{\cal L}/\delta \varphi_I$ is the Euler-Lagrange derivative, and $\simeq$ indicates equality up to a total derivative.  
Expanding~(\ref{invarianceeq}) in powers of $\gamma$ yields a series of equations
\bea \label{symexpan}
\nonumber
&& {\delta^{EL}{\cal L}_{(2)}\over \delta \varphi_I}\delta_{(0)}\varphi_I\simeq 0 \ , \\
&& {\delta^{EL}{\cal L}_{(3)}\over \delta \varphi_I}\delta_{(0)}\varphi_I+ {\delta^{EL}{\cal L}_{(2)}\over \delta \varphi_I}\delta_{(1)}\varphi_I \simeq 0 \ ,\\
\nonumber
&&\vdots 
\eea
For the broken symmetries, we have $\delta_{(0)}\not=0$, and so the first of these equations indicates that $\delta_{(0)}$ is a symmetry of the quadratic action ${\cal L}_{(2)}$.  Note that it is only necessary that $\delta_{(0)}\not=0$ on at least one of the fields.  If there are weight 0 fields or fields with vanishing profiles, the broken symmetries act trivially ($\delta_{(0)}=0$) on these fields but not on the others. (If $\delta_{(0)}=0$ on all the fields, as is the case for the unbroken conformal symmetry, then the second line of (\ref{symexpan}) says that $\delta_{(1)}$ is a symmetry of the quadratic part of the action.)

At the quadratic two derivative level, the most general lagrangian consistent with spatial rotations and translations is\footnote{A term with a single derivative is only possible if that derivative is a time derivative, in which case it can be integrated by parts onto the coefficient and combined with the mass term.} 
\be
{\cal L}_{\rm quad} = \frac{1}{2}M^{IJ}_1(t) \dot{\varphi}_I\dot{\varphi}_J- \frac{1}{2}M^{IJ}_2(t) \vec{\nabla}\varphi_I \vec{\nabla}\varphi_J - \frac{1}{2}M^{IJ}_3(t) \varphi_I\varphi_J\,,
\label{quadgen1}
\ee
where summation over $I, J$ is implicit, and where $M_{\cal I}^{IJ}(t)$, ${\cal I} = 1,2,3$ are symmetric, time-dependent matrices.
Let us start by imposing the linearly-realized symmetries. Since~(\ref{quadgen1}) is manifestly invariant under spatial
rotations and translations, we are left with dilatations $\delta_D$ and the spatial components of the special
conformal transformations $\delta_{K_i}$.

Dilatation invariance imposes the following restrictions on the $M_{\cal I}^{IJ}(t)$'s:
\be
\dot{M}_{1,2}^{IJ}  = \frac{2(d_I - 1)}{t}M_{1,2}^{IJ}\,,\qquad  \dot{M}_3^{IJ}  = \frac{2(d_I - 2)}{t}M_3^{IJ}\,.
\label{Mdil}
\ee
Since these matrices are symmetric, it follows that $0 = \dot{M}_{\cal I}^{IJ} -  \dot{M}_{\cal I}^{JI} = 2 (d_I-d_J) M_{\cal I}^{IJ}/t$,
and hence $M_{\cal I}^{IJ} = 0$ for $d_I\neq d_J$. In other words, fields of different conformal weights do not mix at the quadratic level --- the $M_{\cal I}$'s
are block-diagonal matrices, with each block corresponding to a particular conformal weight. Moreover,~(\ref{Mdil}) fixes the time-dependence
within each block:
\be
M_{1,2}^{IJ}(t) = \tilde{M}_{1,2}^{IJ} (-t)^{2(d_I-1)} \,,\qquad M_{3}^{IJ}(t) = \tilde{M}_{3}^{IJ} (-t)^{2(d_I-2)}\,,
\ee
where the $\tilde{M}_{\cal I}^{IJ}$'s have constant matrix elements. 

Invariance under the spatial components of the special conformal transformations, $\delta_{K_i}$, yields a further restriction
\be
\tilde{M}^{IJ}_2= \tilde{M}^{IJ}_1\,.
\label{M2M1}
\ee
Therefore, the most general form of the lagrangian consistent with the linearly-realized symmetries is
\be
{\cal L}_{\rm quad} =- \frac{1}{2}\tilde{M}^{IJ}_1(-t)^{2(d_I-1)}  \eta^{\mu\nu}\partial_\mu \varphi_I\partial_\nu \varphi_J  - \frac{1}{2}\tilde{M}^{IJ}_3 (-t)^{2(d_I-2)} \varphi_I\varphi_J \,,
\label{quadgen2}
\ee
where the coefficient matrices are block diagonal according to conformal dimension.  Note that the $so(4,1)$ unbroken symmetries imply that perturbations
must propagate at the speed of light.

We now impose the 5 non-linear transformations~(\ref{brokengenmany}) associated
with the broken symmetries. Let us focus on the conformal block of weight $d$, letting the indices $I,J$ run only over this block.  As discussed above, assuming the conformal symmetry is broken ({\it i.e.}, there is at least one field with $d>0$ and non-vanishing profile), we need only impose $\delta_{(0)}$, the non-linear part of the transformations (\ref{brokengenmany}),
\be
\delta_{P_0}\varphi_I = d{ \bar\phi_I \over t} \,, \qquad \delta_{J^{0i}}\varphi_I  =- d{ x^i\over t} \bar\phi_I   \,,\qquad \delta_{K_0}\varphi_I  = -d{ x^2 \over t}\bar\phi_I  .
\label{nonlingenmany}
\ee
Invariance under $\delta_{P_0}$ yields the following condition
\be
d\ \tilde {M}^{IJ}_3c_J = d(d + 1)(d-4)\tilde{M}^{IJ}_1c_J\,,
\label{M3M1}
\ee
where the $c_I$'s are the coefficients of the background solution (\ref{phibackgenmany}).  The other transformations $\delta_{K_0}$ and $\delta_{J^{0i}}$ provide no further constraints.  

By re-defining the fields, we may diagonalize the kinetic matrix within each block, setting $\tilde{M}^{IJ}_1=\delta^{IJ}$ and $\tilde {M}^{IJ}_3\equiv  {M}^{IJ}$.  The condition (\ref{M3M1}) becomes
\be 
d\ {M}^{IJ}c_J= d(d + 1)(d-4)\  c^I \,.
\label{M3M1p}
\ee
For the $d=0$ block, this condition is trivial and we obtain no further conditions on the mass matrix $M^{IJ}$ from the non-linear symmetries.  For the $d\not=0$ blocks, we get an extra condition on $M^{IJ}$, namely that it has $c_I$, the background solution coefficients, as an eigenvector with eigenvalue $(d + 1)(d-4)$.

Hence the most general quadratic lagrangian for fluctuations realizing the desired symmetry breaking pattern is
\be
{\cal L}_{\rm quad} \sim \sum_{\rm blocks} \left(-\frac{1}{2} (-t)^{2(d-1)} \eta^{\mu\nu}\partial_\mu \varphi_I\partial_\nu \varphi^I -  \frac{1}{2} (-t)^{2(d-2)} M^{IJ}\varphi_I\varphi_J \right)\,,
\label{quadfinal}
\ee
up to an overall constant prefactor, and the constraint that for the $d\not=0$ blocks, $M^{IJ}$ has $c_I$ as an eigenvector with eigenvalue $(d+1)(d-4)$. 

More can be said in the case of a $d\neq 0$ block where only one field has non-vanishing background profile. 
Then the mass matrix has only one element, denoted by $M$, and (\ref{M3M1p}) gives  the condition $M=(d + 1)(d-4)$, so that the quadratic action can be written as

\be
{\cal L}_{\rm quad} \sim -\frac{1}{2} (-t)^{2(d-1)} \eta^{\mu\nu}\partial_\mu \varphi\partial_\nu \varphi-  \frac{1}{2} (-t)^{2(d-2)}(d + 1)(d-4)\varphi^2\,\ \ \ ({\rm single\ field, \ non\ vanishing} \ \bar\phi).
\label{quadfinalsingle}
\ee

By imposing the symmetries in a ``brute-force" manner, we have derived the most general quadratic action for the perturbations, up to 2 derivatives. 
In ongoing work, we followed a more systematic approach by applying the well-known coset construction~\cite{Coleman:1969sm,Callan:1969sn} to the 
symmetry breaking pattern of interest~\cite{Hinterbichler:2012mv}. In this way, we can derive the most general effective Lagrangian (including non-linear and higher-derivative terms) that linearly realizes the $so(4,1)$ symmetries and non-linearly realizes the $so(4,2)$ group for any number of fields of arbitrary conformal weight, valid irrespective of the underlying dynamics. Though usually applied to broken internal symmetries, the
construction has been generalized to space-time symmetries~\cite{volkov}. This ongoing work will extend existing results on breaking conformal symmetry down to Poincar\'e~\cite{Isham:1970xz,Salam:1970qk,Isham:1970gz} to the case of breaking to the de Sitter algebra.

\subsection{\label{dynattr}Stability of the Background}

The general quadratic lagrangian allows us to assess the stability of the background~(\ref{phibackgenmany}).  Take for simplicity the case where there is at most one conformal field for each $d\not=0$, and that each has non-vanishing background, so that we may use the action (\ref{quadfinalsingle}) for each field.
Working in Fourier space for the spatial variables, the equation of motion for the mode functions of each field $\varphi$ is
\be
\ddot{\varphi}_k + k^2\varphi_k  + \frac{2(d-1)}{t} \dot{\varphi}_k + \frac{(d + 1)(d-4)}{t^2} \varphi_k = 0\,.
\label{modfcneqn}
\ee
At long wavelengths ($k\rightarrow 0$), or at sufficiently late times such that gradients can be ignored, the growing and decaying mode solutions are respectively given by
\be
\varphi_k \sim \frac{1}{(-t)^{d +1}} \,\,\,\,\,\, {\rm and} \,\,\,\,\, \varphi_k \sim (-t)^{4-d}\,,
\label{growingdecaying}
\ee
where we recall that $-\infty < t < 0$. The growing mode at first sight looks dangerous, since it grows more rapidly than the
background profile~(\ref{phibackgenmany}). 

However, if in the entire theory there is only a single field $\phi$ with non-vanishing profile and weight $d\not=0$,
then this perturbation can be re-summed into a harmless constant time-shift of the background solution:
\be
\bar{\phi}(t + \varepsilon)  = \bar{\phi}(t) +\varepsilon \dot{\bar{\phi}} (t)  \sim  \frac{1}{(-t)^{d}} \left( 1 - \frac{d\varepsilon}{t}\right)\,.
\label{phiattractgen}
\ee
Hence the perturbed field $\phi = \bar{\phi} + \varphi$ tends to the background solution, up to an irrelevant constant shift in time. 
If there are multiple fields with $d\not=0$, on the other hand, then in general each field can be perturbed by a mode (\ref{growingdecaying}) with different independent constant coefficients. Only one of the coefficients can be eliminated with a global time shift, while the rest can be interpreted as time-delays for the various fields. Consequently the solution starts to depart from the desired symmetry breaking solution at late times.\footnote{We thank V.~Rubakov for discussions on this point.}

\subsection{Perturbation Spectrum for Weight $d\neq 0$ Fields}
\label{pertspec}

Next we compute the spectrum of perturbations for a single field with $d\neq 0$ and non-vanishing background, starting from the standard adiabatic vacuum state.
Remarkably, with the field redefinition to the canonical variable $v = (-t)^{d-1}\varphi$, the mode function equation~(\ref{modfcneqn}) reduces to a universal form
\be
\ddot{v}_k  + \left(k^2 - \frac{6}{t^2}\right)v_k = 0\,. \label{modegener}
\ee
The mode function solutions are given by a Hankel function
\be
v_k  \sim \frac{\sqrt{-t}}{4\sqrt{2}\pi}H_{5/2}^{(1)} (-kt) \,,
\label{delphihankelsingle}
\ee
where the ``$\sim$" indicates an unknown normalization constant, tracing back to the overall coefficient in the quadratic action~(\ref{quadfinalsingle}).

In the long wavelength limit, $k|t|\ll 1$, the asymptotic form $H_{5/2}^{(1)}(x) \simeq -3i \sqrt{2/\pi} x^{-5/2}$ implies a strongly red-tilted spectrum,
\be
k^{3/2}|v_k|\sim \frac{3}{4\pi^{3/2}}\frac{1}{kt^2}\,.
\ee
Transforming back to $\varphi = v/(-t)^{d-1}$, this implies
\be
k^{3/2}|\varphi_k | \sim \frac{3}{4\pi^{3/2}}\frac{1}{k(-t)^{d+1}}\,.
\ee
The time-dependence agrees with the growing mode in~(\ref{growingdecaying}), as it should. To summarize, all rolling fields with non-vanishing profile and $d\neq 0$ acquire a universal spectrum for their perturbations. These perturbations are far from scale invariant and cannot be the source of the observed density perturbations. However we will see in Section~\ref{conf0}
that conformal weight 0 fields do acquire scale invariant perturbations on this background, under very general conditions.

The $1/k$ tilt of the Goldstone spectrum is naively worrisome, since it suggests a potential breakdown of perturbation theory on large scales.
When we include mixing with gravity in Section~\ref{gravon}, however, we will find that this in fact translates to a very blue spectrum for the curvature perturbation,
$\zeta_k \sim k/M_{\rm Pl}$, which is therefore under control for all (sub-Planckian) modes of interest. This mismatch in spectra traces back to the fact that $\varphi$ corresponds to
an adiabatic perturbation in our case, unlike in the scenario of~\cite{rub1}.

\subsection{Conformal Weight 0 Modes and Scale Invariance}
\label{conf0}

The source of scale invariant perturbations will be conformal weight 0 fields, which will be denoted by $\chi_I$. 
This field has a trivial background $\bar\chi_I = {\rm const.}$, where the constant can
even be zero. 
The form of the quadratic action is (\ref{quadfinal}) for $d=0$,
\be
{\cal L}_{\rm quad}^{(d = 0)} =- \frac{1}{2} t^{-2}  \eta^{\mu\nu}\partial_\mu \chi_I\partial_\nu \chi^I  - \frac{1}{2}M^{IJ} t^{-4} \chi_I\chi_J\,.
\label{quadgenconf0v2}
\ee
The mass matrix $M^{IJ}$ is unconstrained, so the weight-0 fields generically have mass mixing. 

The action~(\ref{quadgenconf0v2}) is sufficient to compute the spectrum of $\chi$ fluctuations.  
Assuming for simplicity a single field $\chi$, with the field redefinition $\chi = \hat\chi/(-t)$ the mode functions satisfy
\be
\ddot{\hat\chi}_k +\left( k^2  - \frac{2-M}{t^2}\right)\hat\chi_k = 0\,,
\label{fmodefcn}
\ee
where $M$ is the single component of the mass matrix.  From analogous computations in inflation it is well-known that the spectrum will be scale invariant provided that 
$M \ll 1$. Indeed, in this regime, the solution for the mode functions, assuming the standard adiabatic vacuum,
is given by
\be
\hat\chi_k \sim \frac{e^{-ikt}}{\sqrt{2k}}\left(1-\frac{i}{kt}\right)\,,
\label{hatchisol1}
\ee
where once again the mode normalization is not fixed.
The long-wavelength spectrum for $\chi$ is therefore scale invariant,
\be
k^{3/2}|\chi_k| \simeq {\rm constant}\,.
\label{Pchi1}
\ee
Hence, under very general assumptions, a conformal weight 0 field acquires a Harrison-Zeldovich spectrum in
our background~(\ref{phibackgenmany}).

An immediate corollary of the above derivation is that the growing mode solution for $\chi$ is a constant.
Thus perturbations in this field are amplified, but only to a particular finite value. This is a key difference from other non-inflationary multi-field mechanisms.
In the New Ekpyrotic scenario~\cite{Buchbinder:2007ad,Buchbinder:2007tw,Buchbinder:2007at}, for instance, the amplification of scale invariant perturbations in a second field relies on a tachyonic instability~\cite{Koyama:2007mg}. The background solution is therefore unstable to unbounded growth along this field direction, though it was shown in~\cite{Buchbinder:2007tw} that an earlier phase of evolution can
bring the field arbitrarily close to the desired trajectory. Similarly, the general two-field non-inflationary mechanisms of~\cite{Tolley:2007nq} are also tachyonically unstable.

\section{\label{Explicitsection}An Example: Negative Quartic Potential}
\label{phi4}

The simplest realization of our mechanism relies on a canonical scalar field $\phi$ of conformal weight~1 rolling down a {\it negative} $\phi^4$ potential~\cite{rub1}.
The $\phi$ part of the action reads
\be
{\cal L}_\phi = -\frac{1}{2}(\partial\phi)^2 + \frac{\lambda}{4}\phi^4 + \ldots 
\label{Sphi}
\ee
This transforms to a total derivative under~(\ref{delphiconfmany}), hence the theory is conformally invariant. Of course this is only true at the
classical level, and we will discuss radiative corrections shortly. With $\lambda > 0$, corresponding to negative potential energy, it is easy to see from the beta function that this theory is in fact asymptotically free.

\begin{figure} 
   \centering
   \includegraphics[width=5.0in]{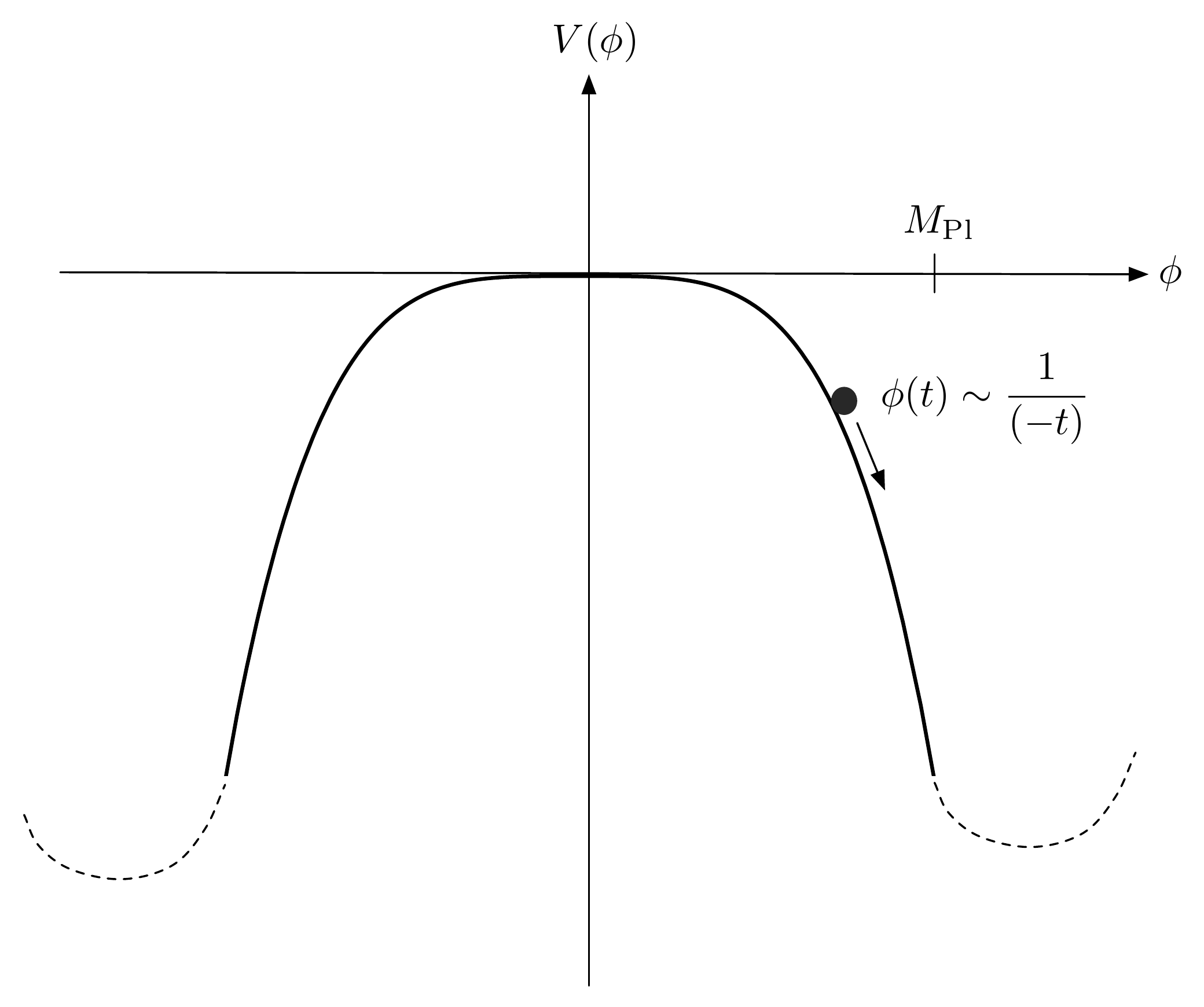}
   \caption{\textit{Sketch of the scalar potential for the simplest realization of our mechanism. The potential is well approximated by a negative
   quartic form along the solid curve. Higher-dimensional operators such as ${\cal O}(\phi^6/M_{\rm Pl}^2)$ become important along the dotted curve,
   stabilizing the potential.}}
   \label{phi4pot}
\end{figure}

Although the potential is unbounded from below as it stands, we envision that higher-dimensional operators --- denoted by the
ellipses --- regularize the potential at large values of $\phi$. Specifically, in Section~\ref{gravon} we will couple this theory minimally to gravity, which will
break the conformal symmetry explicitly at the $1/M_{\rm Pl}$ level. Hence it is natural to expect ${\cal O}(\phi^6/M_{\rm Pl}^2)$ and higher-order
corrections that can turn the potential around. This is shown in Fig.~\ref{phi4pot}.  In fact, we will see in Section~\ref{gravon} that the flat-space approximation in any case breaks down when
$\phi \sim M_{\rm Pl}$. For simplicity, in this Section we work in flat space and take the quartic form as exact.  

This theory allows for a $1/t$ solution of the form~(\ref{phibackgenmany}). Indeed, the equation of motion for $\phi$ in the case of purely time-dependent evolution is
\be
\ddot{\phi} = \lambda\phi^3\,.
\ee
This has a first integral of motion, $E = \dot{\phi}^2 -\lambda\phi^4/4$, which is of course recognized as the total energy density.
The choice of interest is $E = 0$, in which case
\be
\bar\phi = \frac{\sqrt{2}}{\sqrt{\lambda}(-t)}\,.
\label{phiback}
\ee
This describes rolling from $\phi = 0$ asymptotically in the past ($t=-\infty$) towards $\phi \rightarrow \infty$ as $t \rightarrow 0$.
In practice, we will only need a finite portion of this solution to generate a scale invariant spectrum over a sufficiently wide range of scales.
Although we have focused on the $E=0$ solution, as per the discussion of Section~\ref{symmbreak} this solution is a dynamical attractor
for more general initial conditions. 

In the limit that the negative quartic form is exact, the theory is conformally invariant, at least
at the classical level. Of course quantum corrections break the conformal symmetry, \`a la
Coleman-Weinberg~\cite{Coleman:1973jx}. The 1-loop effective potential is given by~\cite{Weinberg:1996kr,Weinberg:1987vp}
\be
V_{\rm CW}(\phi) =  - \frac{4\pi^2\phi^4}{9\ln(\phi^2/\bar{M}^2)}\,,
\label{VCW}
\ee
where $\bar{M}$ is the mass scale introduced during renormalization\footnote{Equation~(\ref{VCW}) obtains by integrating over
non-tachyonic modes only, {\it i.e.} with $k^2 - 3\lambda\phi^2 > 0$. Integrating over all momentum shells
would result in a complex potential, whose imaginary part encodes the decay rate from an initially homogeneous field profile
to domains of different field values~\cite{Weinberg:1987vp}. Here we instead only integrate out non-tachyonic modes and describe tachyonic modes
as perturbations over our classical solution.}. 
In the Coleman-Weinberg potential~(\ref{VCW}), the quartic coupling $\lambda(\phi)$ is field-dependent. If this coupling is small
for some fiducial field value $\phi_{\rm fid}$,
\be
\lambda(\phi_{\rm fid}) \ll 1\,,
\ee
then $\lambda(\phi)$ will be even smaller, and hence the 1-loop potential~(\ref{VCW}) increasingly accurate, for $|\phi|$ values
increasingly larger than $|\phi_{\rm fid}|$. In other words, perturbation theory can be trusted for arbitrarily large field values.
In our case, we will see in Section~\ref{gravon} that in the presence of gravity our approximations are valid only for
$\phi\,\gsim\, \phi_{\rm beg} = \lambda^{1/4}M_{\rm Pl}$. Hence it is natural to take $\phi_{\rm fid} = \phi_{\rm beg}$,
thereby ensuring that the quartic coupling is small over the entire evolution. With this definition of $\lambda$, we
henceforth approximate the effective potential as a pure quartic form and neglect the small logarithmic corrections.

In addition to the zero-derivative corrections encoded in the Coleman-Weinberg potential, our background solution is robust against generic higher derivative corrections that are invariant under the assumed conformal symmetries.
These are of the general form
\be
\frac{\partial^{2n}\phi^m}{\phi^{2n+m-4}}\,,
\label{gencor}
\ee
where $n,m = 0,1,2\ldots$, and where the $n$ derivatives can act on any number of the $m$ fields. 
Terms of the form~(\ref{gencor}) are manifestly invariant under dilatations\footnote{They are not generally invariant under the special conformal transformations, though suitable linear combinations of them may be.  At $n=2$, for instance, we have the conformal galileon combination~\cite{Nicolis:2008in}
\be
\frac{1}{\phi^3} (\partial\phi)^2\Box\phi - \frac{1}{2\phi^4}(\partial\phi)^4\,.
\ee
 In any case, if the scale invariant terms ~(\ref{gencor}) are unimportant, then the more restrictive subset of conformally invariant terms will be unimportant.}.
For $n=0$, the only allowed term is $\phi^4$, which we have already considered.
For $n=1$, all terms included in~(\ref{gencor}) are equivalent by integration by parts to $(\partial\phi)^2$.
Hence we assume $n\geq 2$ without loss of generality. 
On our background solution~(\ref{phiback}), $\dot{\bar{\phi}}\sim \bar{\phi}/t$, $\ddot{\bar{\phi}} \sim \bar{\phi}/t^2$, etc.
In other words, each derivative introduces one power of $1/t$, hence the general term~(\ref{gencor}) scales as
\be
\frac{\partial^{2n}\phi^m}{\phi^{2n+m-4}} \sim \frac{1}{t^{2n}\phi^{2n-4}} \sim \lambda^n\phi^4\,.
\ee
Since $n \geq 2$, these terms are all suppressed by additional powers of $\lambda$ compared to the quartic term $\lambda\phi^4$.
Therefore, even if we allow $O(1)$ coefficients for generic corrections~(\ref{gencor}), we are justified in neglecting
them to derive our background solution.

\subsection{Perturbations in the Rolling Field}
\label{phipertphi4}

Expanding~(\ref{Sphi}) in terms of the Goldstone perturbation $\varphi = \phi -\bar\phi$, we obtain
\be
{\cal L}_\varphi = -\frac{1}{2}(\partial\varphi)^2 + \frac{3}{2}\lambda\bar\phi^2 \varphi^2 + \lambda\bar{\phi} \varphi^3 +  \frac{\lambda}{4}  \varphi^4\,.
\ee
At quadratic order, this is consistent with the general result~(\ref{quadfinalsingle}) with $d = 1$. 
(The unknown overall constant prefactor in~(\ref{quadfinalsingle}) is fixed to unity, since the kinetic term is
canonical for the case at hand.) The mode functions $\varphi_k$ satisfy
\be
\ddot{\varphi}_k + k^2\varphi_k = \frac{6}{t^2}\varphi_k\,,
\label{modefcnspecific}
\ee
in agreement with~(\ref{modegener}). (For $d = 1$, the field redefinition to the $v$ variable is trivial: $v = \varphi$.) As discussed in Section~\ref{dynattr}, this implies that the $1/t$ background is a dynamical attractor. Assuming the standard adiabatic vacuum,~(\ref{modefcnspecific}) can be solved exactly in terms of Hankel functions 
\be
\varphi_k = \frac{\sqrt{-t}}{4\sqrt{2}\pi}H_{5/2}^{(1)} (-kt) \,.
\label{delphihankel}
\ee
This matches~(\ref{delphihankelsingle}), except that the prefactor has now been fixed by the canonical normalization assumed here.
The Goldstone spectrum at long wavelengths is therefore given by
\be
k^{3/2}|\varphi_k|\simeq \frac{3}{4\pi^{3/2}}\frac{1}{kt^2}\,.
\label{delphigrows}
\ee
Hence each mode corresponds to a time delay
\be
\frac{k^{3/2}|\delta t_k|}{|t|} = \frac{k^{3/2}|\varphi_k|}{t\dot{\bar{\phi}}} = \frac{3\sqrt{\lambda}}{4\sqrt{2}\pi^{3/2}k|t|}\,.
\ee
We will see in that Section~\ref{gravon} the time-delay mode actually projects out of the curvature perturbation, $\zeta$,
resulting instead in a strongly blue-tilted adiabatic spectrum: $\zeta_k \sim k/M_{\rm Pl}$. 

Conservatively, one can demand that $k^{3/2}|\delta t_k|/|t|$ be small for the relevant range of modes by making $\lambda$ sufficiently small.
For a given value of $\lambda$, it is clear that perturbation theory holds for a finite time interval $t_{\rm i} < t < t_{\rm f}$.
Since modes ``freeze-out" when $k|t| = 1$, let us denote by  $k_{\rm i} \equiv 1/|t_{\rm i}|$ and $k_{\rm f}\equiv 1/|t_{\rm f}|$
the respective wavenumbers. Demanding that perturbation theory be valid over this range of modes requires
\be
\lambda \, \lsim\, \frac{k_{\rm i}^2}{k_{\rm f}^2} \equiv e^{-2N}\,,
\label{lambdaboundnaive}
\ee
where $N$ is the corresponding number of e-folds of perturbations. It has been argued in~\cite{rub3} that large time-delays
may be sensible, in which case~(\ref{lambdaboundnaive}) would be unnecessary. In any case, other considerations in Section~\ref{gravon} will limit the range of modes for a given $\lambda$ 
that can be produced within the regime of validity of our approximations, resulting in the much tighter constraint $\lambda \sim e^{-4N}$.

\subsection{Scale Invariant Weight-0 Perturbations}
\label{pert}

As shown in Section~\ref{conf0}, scale invariant perturbations originate from a conformal weight 0 field $\chi$, so let us now add such a field to the lagrangian. The most general, conformally-invariant lagrangian we can add to~(\ref{Sphi}) with at most 2 derivatives and involving only $\chi$ and $\phi$ is
\be
{\cal L}_\chi = - \frac{1}{2} \phi^2 \left(\partial\chi\right)^2 - \kappa \lambda \phi^4V(\chi) + \xi \phi\Box\phi F(\chi) \,,
\label{Lchi}
\ee
where $\kappa$ and $\xi$ are dimensionless constants. This transforms to a total derivative under the transformations~(\ref{delphiconfmany}), hence the theory is conformal at the classical level. (Note that a term of the form $(\partial\phi)^2 G(\chi)$ is invariant under dilatations but breaks special conformal symmetries unless $G = {\rm constant}$, in which case it
can be absorbed in a redefinition of $\lambda$.) The kinetic term could be multiplied by an arbitrary function $K(\chi)$, but a field redefinition of $\chi$ would bring the lagrangian
back to the canonical form~(\ref{Lchi}).

Let us assume that $V(\chi)$ and $F(\chi)$ can be Taylor-expanded around $\chi = 0$:
\bea
\nonumber
V(\chi) &=& V\vert_{\chi=0} + \frac{\partial V}{\partial\chi}\bigg\vert_{\chi = 0} \chi + \frac{1}{2}\, \frac{\partial^2 V}{\partial\chi^2}\bigg\vert_{\chi = 0} \chi^2 + \ldots  \\
F(\chi) &=& F\vert_{\chi=0} + \frac{\partial F}{\partial\chi}\bigg\vert_{\chi = 0} \chi + \frac{1}{2}\, \frac{\partial^2 F}{\partial\chi^2}\bigg\vert_{\chi = 0} \chi^2 + \ldots 
\eea
The constants $V\vert_{\chi=0}$ and $F\vert_{\chi=0}$ can be absorbed in a redefinition of $\lambda$ and a wavefunction renormalization, respectively, so we set these terms
to zero without loss of generality.
Moreover, we assume that $\partial V/\partial\chi\vert_{\chi = 0} = \partial F/\partial\chi\vert_{\chi = 0} = 0$, so that $\chi = 0$ is a solution. Finally, through rescaling of $\kappa$ and $\xi$,
we can assume $\partial^2 V/\partial\chi^2\vert_{\chi = 0} = \partial^2 F/\partial\chi^2\vert_{\chi = 0} = 1$. Therefore, at quadratic order in $\chi$, the lagrangian takes the general form
\be
{\cal L}_\chi = - \frac{1}{2} \phi^2 \left(\partial\chi\right)^2 - \frac{\kappa}{2} \lambda\phi^4\chi^2 + \frac{\xi}{2} \phi\Box\phi \chi^2 + {\cal O}(\chi^3) \,.
\label{Lchigen}
\ee
We will now use this lagrangian to derive the power spectrum for $\chi$ generated on the symmetry-breaking background $\bar\phi(t)$.

Since $\chi = 0$ on the background, this gives, at quadratic order in perturbations,
\be
{\cal L}_\chi^{\rm quad}  =  - \frac{1}{2} \bar\phi^2 \left(\partial\chi\right)^2 - \frac{\kappa}{2} \lambda\bar\phi^4\chi^2 +  \frac{\xi}{2} \bar\phi\Box\bar\phi \chi^2 \,.
\ee
Redefining the field in terms of the canonically-normalized variable
\be
\hat\chi \equiv \bar{\phi}\, \chi\,,
\label{chihat}
\ee
and substituting the background solution~(\ref{phiback}), the quadratic lagrangian simplifies to
\be
{\cal L}_{\hat\chi}^{\rm quad}  =  - \frac{1}{2} \left(\partial \hat\chi \right)^2  + \frac{2(1 - \kappa-\xi)}{t^2} \frac{\hat\chi^2}{2}\,.
\ee
This field therefore has a time-dependent mass, exactly as if it were living on a cosmological background. But the true background is of course flat,
Minkowski space, and the time-dependent mass originates from the background evolution of $\phi$.

It follows that the mode functions $ \hat\chi_k$ satisfy
\be
\ddot{\hat\chi}_k + k^2\hat\chi_k   - \frac{2(1 - \kappa-\xi)}{t^2} \hat\chi_k = 0\,.
\label{hatchieom}
\ee
This is exactly of the form~(\ref{fmodefcn}), with $M = 2(\kappa + \xi)$. The general solution is given as usual in
terms of Hankel functions. It is well-known from cosmological perturbation theory that the resulting spectrum will be
scale invariant provided that $|\kappa|, |\xi| \ll 1$. Indeed, in the limit that the $\kappa$ and $\xi$ terms are negligible,
the mode functions are given by
\be
\hat\chi_k \simeq \frac{e^{-ikt}}{\sqrt{2k}}\left(1-\frac{i}{kt}\right)\,,
\ee
where the standard adiabatic vacuum choice has been made. This is consistent with~(\ref{hatchisol1}). In the terms of the original variable $\chi$, this implies
\be
\chi_k =  \frac{\sqrt{\lambda}(-t) e^{-ikt}}{2\sqrt{k}}\left(1-\frac{i}{kt}\right)\,.
\ee
In the long wavelength ($k|t|\ll 1$) limit, the amplitude is indeed scale invariant:
\be
k^{3/2}|\chi_k| = \frac{\sqrt{\lambda}}{2} \,.
\label{chiflatfinal}
\ee
This agrees with the general result~(\ref{Pchi1}), except that the normalization of the two-point function has now been fixed in terms
of the parameters of the theory under consideration. 

Meanwhile, the spectral tilt can be immediately inferred from~(\ref{hatchieom}):
\be
\left(n_s - 1\right)_{\kappa,\xi} = \frac{4}{3}(\kappa + \xi)\,.
\ee
This contribution to the spectral index is independent of scale. The spectrum is nearly scale invariant for $|\kappa|,|\xi| \ll 1$,
and has a red (blue) tilt for $\kappa + \xi < 0$ ($>0$).

\subsection{Validity of the Effective Theory}
\label{loops}

Next we would like to assess the validity of the effective theory obtained by perturbing around the symmetry-breaking background $\bar\phi(t)$.
In particular, since it is already obvious from~(\ref{lambdaboundnaive}) that we will be interested in exponentially small values of $\lambda$, an immediate concern
is whether this is stable or not under quantum corrections. For the purpose of estimating radiative corrections we will use power counting arguments, using the lowest scale suppressing
higher-dimensional operators for cutting off loop integrals\footnote{In other words, we do not make optimistic assumptions about the UV completion, as is implicitly done, for instance, when
using dimensional regularization with minimal subtraction.}.

Letting $\phi = \bar{\phi}(t) + \varphi(t,\vec{x})$ and $\chi = \chi(t,\vec{x})$, the lagrangian given by the sum of~(\ref{Sphi}) and~(\ref{Lchigen}) can be written as, up to quadratic order in $\chi$,
\bea
\nonumber
{\cal L} &=&  -\frac{1}{2}(\partial\varphi)^2 + \frac{3}{2}\lambda \bar{\phi}^2\varphi^2 + \lambda\bar{\phi}\varphi^3 + \frac{\lambda}{4}\varphi^4  \\
\nonumber
& & -\frac{1}{2}\left(\bar{\phi}^2 + 2\bar{\phi}\varphi + \varphi^2\right)(\partial\chi)^2 - \frac{\kappa}{2}\lambda \left(\bar{\phi}^4 + 4\bar{\phi} \varphi^3   + 4\bar{\phi}^3 \varphi + 6\bar{\phi}^2\varphi^2  + \varphi^4 \right)\chi^2 \\
& & + \frac{\xi}{2}\left(\bar\phi\Box\bar\phi + \varphi\Box\bar\phi +\bar\phi\Box\varphi +\varphi\Box\varphi\right)\chi^2\,.
\label{Lperturb1st}
\eea
For sufficiently small space-time regions, $\bar{\phi}$ can be treated as a constant in this lagrangian. Translating to the canonically-normalized variable $\hat\chi = \bar\phi \chi$ introduced in~(\ref{chihat}),
the perturbed lagrangian becomes
\bea
\nonumber
{\cal L} &=& -\frac{1}{2}(\partial\varphi)^2 + \frac{3}{2}\lambda \bar{\phi}^2\varphi^2 + \lambda\bar{\phi}\varphi^3 + \frac{\lambda}{4}\varphi^4  \\
\nonumber
&-& \frac{1}{2}(\partial\hat{\chi})^2 - \frac{\kappa}{2}\lambda \bar{\phi}^2\hat{\chi}^2 - 2\kappa \lambda \bar{\phi}\hat{\chi}^2\varphi  -  3\kappa \lambda \hat{\chi}^2\varphi^2\\
&-& \frac{\varphi}{\bar{\phi}} (\partial\hat{\chi})^2 - \frac{\varphi^2}{2\bar{\phi}^2}(\partial\hat{\chi})^2 - \frac{2\kappa \lambda}{\bar{\phi}}\hat{\chi}^2\varphi^3 - \frac{\kappa\lambda}{2\bar{\phi}^2}\hat{\chi}^2\varphi^4  + \frac{\xi}{2\bar\phi}\hat\chi^2\Box\varphi + \frac{\xi}{2\bar\phi^2}\hat\chi^2 \varphi\Box\varphi  \,.
\label{Lpert}
\eea
The first and second lines include all renormalizable interactions. In particular, the last term in each of the first two lines involves the dimensionless couplings $\lambda$ and $\kappa\lambda$, respectively, which we already know must satisfy $\lambda, |\kappa| \ll 1$ in order to generate a nearly scale invariant spectrum spanning many e-folds of perturbations. Hence these interactions are very weak. 

Meanwhile, the last line in~(\ref{Lpert}) includes higher-dimensional operators. Evidently, these are suppressed by the following scales: $\bar{\phi}$, $\bar{\phi}/\sqrt{|\kappa| \lambda}$, $\bar{\phi}/|\kappa| \lambda$, $\bar\phi/|\xi|$ and $\bar\phi/\sqrt{|\xi|}$.  Since $|\kappa| \lambda, |\xi| \ll 1$, the smallest of these scales is $\bar{\phi}$, which naively is the cutoff of this theory.  However, this cutoff is not physical --- it is merely an artifact of the parametrization of the fields.  This can be seen, for instance by writing the theory in terms of a complex field $\Phi=\phi e^{i\chi}$.  For $\kappa=\xi =0$, the theory becomes the theory of a canonical complex field with a $|\Phi|^4$ potential, which has no $\bar\phi$ cutoff.  Alternatively, one can compute a quantity independent of field redefinitions, for instance the on-shell amplitude for $\chi\chi\rightarrow\chi\chi$ scattering using the naively non-renormalizable interaction $\sim\varphi (\partial\chi)^2/\bar\phi$.  The amplitude at energies $\gg m_\varphi \sim\sqrt{\lambda}\bar\phi$ is proportional to $\sim \bar\phi^{-2} \left(s+t+u\right)\sim m_\chi^2/ \bar\phi^2$, and hence does not violate unitarity at energies of order $\bar\phi$.  

Thus a conservative estimate of the strong coupling scale of the theory is
\be 
\Lambda_s=\min \left({\bar{\phi}\over \sqrt{|\kappa|\lambda}},\frac{\bar\phi}{\sqrt{|\xi|}}\right) \ \ \ \ ({\rm without}\,\,{\rm gravity})\,. 
\label{strongcoup}
\ee
Not surprisingly, the cutoff of the theory is set by the background solution and hence is time-dependent. The cutoff is small initially and grows monotonically in time.

The perturbation lagrangian~(\ref{Lpert}) was derived under the approximation that $\bar\phi$ is approximately constant. In other words, gradients and time derivatives in~(\ref{Lpert})
are assumed large compared to the characteristic frequency associated with the background: $|\vec\nabla|,|\partial_t| \gg \omega \equiv \dot{\bar{\phi}}/{\bar{\phi}}= 1/(-t)$.
On the other hand, the validity of the effective theory requires $|\vec\nabla|,|\partial_t| \ll \Lambda_s$. Therefore, for consistency, the characteristic frequency must also be well below the cutoff:
\be
\frac{\omega}{\Lambda_s} = \frac{\dot{\bar{\phi}}}{\Lambda_s \bar{\phi} } = \max\left( \frac{ \sqrt{|\kappa|\lambda}}{\bar{\phi} (-t)}, \frac{ \sqrt{|\xi|}}{\bar{\phi} (-t)}\right) \sim \max\left ( \sqrt{|\kappa|}\lambda, \sqrt{|\xi|\lambda}\right) \ll 1\,. 
\label{flatspacecheck}
\ee
Happily, there is indeed a large parametric window for which $\bar\phi$ can be treated as approximately constant while gradients and time derivatives of perturbations are well below the cutoff.

Using $\Lambda_s $ to estimate various loop diagrams, it is immediately clear that the parameters of~(\ref{Lpert}) will
receive large quantum corrections. For instance, consider the sample 1-loop contributions to the $\lambda \varphi^4$ vertex shown at the top of Fig.~\ref{loopcontributions}.
These diagrams are dangerous because they renormalize $\lambda$ without themselves involving $\lambda$ vertices. Indeed, power counting gives
\be
\delta\lambda \sim \frac{\Lambda_s^4}{\bar\phi^4} \,,
\label{dellamb1}
\ee
which combined with~(\ref{strongcoup}) implies large corrections to $\lambda$, thereby destabilizing the desired regime $\lambda \ll 1$.
However, when gravity is turned on in Section~\ref{gravon}, we will see that the cutoff of the theory is considerably lowered,
\be
\Lambda_s \, \lsim\, \lambda^{1/4}\bar\phi\qquad \,\,\,\,\, ({\rm with}\,\,{\rm gravity})\,.
\ee
Therefore~(\ref{dellamb1}) instead gives $\delta\lambda \, \lsim\, {\cal O}(\lambda)$, hence $\lambda\ll 1$ is stable under quantum corrections in this case.
We will come back to estimating radiative corrections in Section~\ref{loopestimates}, but first we turn to the description of the scenario in the presence of gravity.

\section{Turning on Gravity}
\label{gravon}

The simplest way to include dynamical gravity is to couple the fields in our CFT minimally to Einstein gravity. As a result, the conformal symmetry is broken explicitly
by $M_{\rm Pl}$-suppressed operators. For concreteness, in this Section we focus on the negative quartic example, though most of the results
will be completely general. The covariant action of interest is given by\footnote{As mentioned in the Introduction, at this point we diverge from the scenario proposed by~\cite{rub1}. Instead of assuming a conformal coupling to the metric, we consider minimal coupling to gravity. Moreover, instead of $\phi$ being a spectator field, as assumed in~\cite{rub1}, we take $\phi$ to drive the cosmological background.}
\be
S = \int {\rm d}^4 x\sqrt{-g}\left( \frac{M_{\rm Pl}^2}{2} R - \frac{1}{2}g^{\mu\nu}\partial_\mu\phi\partial_\nu\phi + \frac{\lambda}{4}\phi^4\right) \,.
\label{Sein}
\ee
Similarly, we can straightforwardly covariantize the lagrangian~(\ref{Lchigen}) for $\chi$:
\be
S_\chi = \int {\rm d}^4 x \sqrt{-g} \bigg(-\frac{1}{2}\phi^2 g^{\mu\nu} \partial_\mu\chi\partial_\nu\chi  -\frac{\kappa}{2}\lambda \phi^4 \chi^2 + \frac{\xi}{2} \phi\Box\phi \chi^2  +  {\cal O}(\chi^3) \bigg)\,.
\label{Schigrav}
\ee
Using these, we can derive the cosmological evolution, under the assumptions of homogeneity, isotropy and spatial flatness: 
${\rm d}s^2 = -{\rm d}t^2 + a^2(t){\rm d}\vec{x}^2$. Since the dynamics are intrinsically non-inflationary --- the scalar potential is negative --- one
may question whether these assumptions are justified. But we will in fact argue in Section~\ref{flatness} that, remarkably, for more general initial conditions
the universe is driven towards increasing degree of homogeneity, isotropy and spatial flatness. 

We assume that the background is driven by $\phi$, with $\chi$ playing the role of a spectator field. The equation of motion for $\phi$ is given by
\be
\ddot{\phi} + 3H \dot{\phi} = \lambda \phi^3\,.
\label{phieomcosmo}
\ee
At sufficiently early times, it turns out that the effects of gravity are negligible, and the flat-space description is a good approximation to the scalar field dynamics. 
We will check {\it a posteriori} that this is indeed the case and identify the time up to which neglecting gravity is a valid approximation. With this assumption, we can
neglect the Hubble damping term and obtain~(\ref{phiback}) as an approximate solution,
\be
\phi(t) \simeq \frac{\sqrt{2}}{\sqrt{\lambda}(-t)}\,.
\label{phiback2}
\ee
(Here, and for most of this Section, $\phi(t)$ implicitly refers to the background solution, and we omit the ``bar" to simplify notation.)

At this order of approximation, the solution~(\ref{phiback2}) has zero energy density, hence the Friedmann equation implies vanishing Hubble parameter.
To solve for $H(t)$, we must therefore go to next order and derive corrections to~(\ref{phiback2}). Alternatively, we can solve the $\dot{H}$ equation,
\be
\dot{H} = -\frac{1}{2M_{\rm Pl}^2}\dot{\phi}^2 = -\frac{1}{\lambda t^4M_{\rm Pl}^2}\,,
\label{dotH}
\ee
to obtain
\be
H(t) = \frac{1}{3\lambda t^3M_{\rm Pl}^2}\,.
\label{HE}
\ee
Since $-\infty < t < 0$, it follows that $H < 0$: the universe is {\it contracting}.  

We are now in a position to check the validity of neglecting gravity in the evolution for $\phi$. Going back to~(\ref{phieomcosmo}),
this amounted to the approximation $|\ddot{\phi}| \gg |3H \dot{\phi}|$, or using~(\ref{HE}),
\be
\left\vert \frac{\ddot{\phi}}{3H\dot{\phi}}\right\vert = 2\lambda t^2M_{\rm Pl}^2\gg 1\,.
\ee
Hence our approximation is valid as long as $t \ll t_{\rm end}$, where 
\be
t_{\rm end} \sim -\frac{1}{\sqrt{\lambda}M_{\rm Pl}}\,.
\label{tend}
\ee
From~(\ref{phiback2}), this corresponds to 
\be
\phi_{\rm end} \sim M_{\rm Pl}\,.
\label{phiend}
\ee
Thus the $1/t$ solution for $\phi$ is a good approximation until $\phi$ reaches values $\sim M_{\rm Pl}$. This is in any case where
we expect higher-dimensional operators to become important and alter the form of the quartic potential, as sketched in Fig.~\ref{phi4pot}. 

The scalar field energy density follows immediately from the Friedmann equation:
\be
\rho_\phi \equiv \frac{1}{2}\dot{\phi}^2 + V(\phi) = 3H^2M_{\rm Pl}^2 = \frac{1}{3\lambda^2t^6M_{\rm Pl}^2}\,.
\ee
Meanwhile, the pressure can be read off from~(\ref{dotH}) using $\dot{H} = -(\rho_\phi + P_\phi)/2M_{\rm Pl}^2$, that is,
$P_\phi = -(2M_{\rm Pl}^2\dot{H} + \rho_\phi)$. But since $\dot{H} \sim 1/t^4$ and $\rho_\phi \sim 1/t^6$ scale
as different powers of time, the $\dot{H}$ term gives the dominant contribution in this expression for $P_\phi$ when $t \ll t_{\rm end}$. Thus,
\be
P_\phi \equiv \frac{1}{2}\dot{\phi}^2 - V = \frac{2}{\lambda t^4}\,.
\ee
The corresponding equation of state is given by 
\be
w = \frac{P_\phi}{\rho_\phi} = 6\lambda t^2M_{\rm Pl}^2\,.
\ee
In other words, the scalar field behaves as a very stiff fluid, $w\gg 1$, for $t \ll t_{\rm end}$. Over the range $-\infty < t < t_{\rm end}$,
the equation of state decreases from $+\infty$ to a value of ${\cal O}(1)$. A contracting phase with $w \gg 1$ is characteristic of
ekpyrotic cosmologies. The key difference here compared to earlier ekpyrotic scenarios is that $w$ decreases rapidly in time, as opposed to
being nearly constant~\cite{Khoury:2001wf} or growing rapidly~\cite{Khoury:2009my,Khoury:2011ii}.

The scale factor is slowly contracting in the regime $t\ll t_{\rm end}$:
\be
a(t) \simeq 1 - \frac{1}{6\lambda t^2M_{\rm Pl}^2}\,.
\ee
Thus we have $a \simeq 1$ throughout the evolution, consistent with the assumption that gravity is a negligible effect
on the scalar field dynamics. It is well known that a contracting universe with $w\gg 1$ is a dynamical attractor~\cite{Gratton:2003pe,Creminelli:2004jg,Erickson:2003zm,Garfinkle:2008ei}, and we will confirm this fact with perturbation theory in Section~\ref{2pt}. Such a phase also drives the universe to be increasingly isotropic, flat and empty,
as we now discuss.

\subsection{Flatness, Homogeneity and Isotropy Problems}
\label{flatness}

Though our background was derived under a set of seemingly unjustified assumptions, namely homogeneity, isotropy, flatness and $\phi$-domination,
we can now argue that this solution is in fact an attractor for a broad range of initial conditions. Consider the general Friedmann equation,
\be
3H^2 M_{\rm Pl}^2 = - \frac{3K}{a^2} + \frac{C_{\rm mat}}{a^3} + \frac{C_{\rm rad}}{a^4} + \frac{C_{\rm aniso}}{a^6}  + \ldots + \rho_\phi\,,
\ee
which includes contributions from spatial curvature ($\sim 1/a^2$), non-relativistic ($\sim 1/a^3$) and relativistic ($\sim 1/a^4$) species, and anisotropic stress
or kinetic energy ($\sim 1/a^6$). The key observation is that since $a\simeq 1$ throughout the evolution, the energy density in each of these components
remains essentially constant. Meanwhile, the energy density in the scalar field increases rapidly in time, $\rho_\phi \sim 1/t^6$, and therefore grows by many orders of
magnitude during the evolution. Thus, even with comparable initial density fractions for the various components, the universe quickly becomes flat, homogeneous, isotropic and empty. Our cosmological background is therefore a dynamical attractor.

To illustrate this more rigorously, consider the spatial curvature component. Since the scale factor is nearly constant for our solution, the fractional curvature
contribution to the Friedmann equation satisfies
\be
\Omega_{\rm K} \equiv \frac{1}{a^2H^2} \simeq \frac{1}{H^2}\,.
\ee
Since $|H|$ increases in time, the universe is driven towards spatial flatness. Indeed, assuming the universe starts out
with $\Omega_{\rm K}^{(i)}\, \lsim \, {\cal O}(1)$ at some initial time $t_{\rm i}$, the final value of $\Omega_{\rm k}^{\rm end}$ at $t_{\rm end}$ is therefore
\be
\Omega_{\rm K}^{\rm end} \sim \frac{H^2(t_{\rm i})}{H^2(t_{\rm end})} = \left(\frac{t_{\rm end}}{t_{\rm i}}\right)^6 = \left(\frac{1}{\sqrt{\lambda}M_{\rm Pl}|t_{\rm i}|}\right)^6\,.
\ee
Clearly, this can be made arbitrarily small by choosing $|t_{\rm i}|$ suitably large.

\subsection{Power Spectra for Adiabatic and Entropy Perturbations}
\label{2pt}

In this Section we turn to the analysis of cosmological perturbations and compute the power spectra for adiabatic and entropy perturbations.
Starting with the adiabatic mode, we saw in Section~\ref{phipertphi4} that in the absence of gravity the fluctuations in $\phi$ acquired a strongly
red spectrum --- see~(\ref{delphigrows}). We will see that with gravity this in fact corresponds to a very blue spectrum for the curvature
perturbation $\zeta$. This is analogous to what was found in the original (single-field) ekpyrotic scenario, where the scale invariant
time-delay mode projects out of $\zeta$, resulting in a strongly blue-tilted adiabatic spectrum~\cite{Khoury:2001zk,Lyth:2001pf,Brandenberger:2001bs,Creminelli:2004jg}.

To study perturbations, let us work in comoving gauge, defined by $\delta\phi= 0$ and $h_{ij} = a^2e^{2\zeta} \delta_{ij}$, which completely fixes the gauge.
The quadratic action for $\zeta$ is given by
\be
S_{\rm quad} = M_{\rm Pl}^2 \int {\rm d}^3x{\rm d}\tau a^2 \epsilon \left[\zeta'^2 - (\vec{\nabla}\zeta)^2\right]\,,
\ee
where primes denote differentiation with respect to conformal time $\tau$. The equation of state parameter $\epsilon$ is defined by
\be
\epsilon = -\frac{\dot{H}}{H^2} = 9\lambda t^2M_{\rm Pl}^2\,,
\label{epsilonE}
\ee
where we have used~(\ref{dotH}) and~(\ref{HE}). In terms of the canonically normalized variable $v = M_{\rm Pl}a\sqrt{2\epsilon} \zeta$,
we have
\be
S_{\rm quad} = \frac{1}{2}\int {\rm d}^3x{\rm d}\tau \left[v'^2 - (\vec{\nabla}v)^2 -\frac{5}{3\lambda t^4 M_{\rm Pl}^2} v^2\right]\,.
\ee
As usual, this describes a scalar field on flat space with a time-dependent mass. (Note that to obtain the mass term
one must consistently work to subleading order in $\lambda^{-1} t^{-2}M_{\rm Pl}^{-2}\ll 1$.)

The mode function equation is therefore given by
\be
v_k'' + \left(k^2 +  \frac{5}{3\lambda t^4 M_{\rm Pl}^2} \right)v_k = 0\,.
\label{vmode}
\ee
With adiabatic vacuum initial conditions, an approximate expression for the growing mode solution is
\be
v_k \simeq \frac{e^{-ikt}}{\sqrt{2k}} \left(1 -  \frac{5}{18\lambda t^2M_{\rm Pl}^2} + \ldots \right)\,,
\ee
where the second term is a small correction for $t \ll t_{\rm end}$. This approximate form captures the mode function behavior
both in the short-wavelength ($k^2 \gg \lambda^{-1}t^{-4}M_{\rm Pl}^{-2}$) and long-wavelength ($k^2 \ll \lambda^{-1}t^{-4}M_{\rm Pl}^{-2}$)
limits. Thus $v_k$ approximately maintains the spectral dependence of the vacuum state. To leading order, therefore, $\zeta$ is given by
\be
k^{3/2}|\zeta_k| = \frac{k^{3/2}|v_k|}{M_{\rm Pl}a\sqrt{2\epsilon} } \simeq \frac{k}{6M_{\rm Pl}^2\sqrt{\lambda}(-t)}\,.
\label{zetafinal}
\ee
Thus the spectrum for the adiabatic mode is strongly blue-tilted, as claimed, and consequently its amplitude is tiny on large scales.

Although the $1/t$ growth of $\zeta$ naively suggests an instability of the background solution, this is not so. As discussed in detail
in~\cite{galgen}, this mode in fact represents a small time-shift of the background $a(t)$ solution.
This is consistent with the flat space analysis of Section~\ref{dynattr} in terms of $\varphi$ and confirms
that our background remains a dynamical attractor in the presence of gravity.

That said, it is important to check that this growth remains within the regime of perturbation theory. This is indeed the case.
Even by the time $t_{\rm end} = -1/\sqrt{2\lambda}M_{\rm Pl}$, when the quasi-static approximation breaks down, we have
\be
k^{3/2}|\zeta_k|_{t = t_{\rm end}} \sim k/M_{\rm Pl}\,, 
\label{zetaampfinal}
\ee
which is $\ll 1$ since all modes of interest are sub-Planckian. Hence the spectrum
for $\zeta$ remains perturbative for all times. For similar reasons, it is easily checked that the 3-point function
for the adiabatic mode is suppressed.

We now turn our attention to the spectrum of entropy perturbations, encoded in the spectator field $\chi$.
Since the background is slowly contracting, we can treat the perturbations as living on flat space, as in Sec.~\ref{pert}.
Thus $\chi$ acquires a scale invariant spectrum with amplitude~(\ref{chiflatfinal}):
\be
k^{3/2}|\chi_k| = \frac{\sqrt{\lambda}}{2} \,.
\label{chiampfinal}
\ee
This spectrum of entropy perturbations can be converted to the adiabatic mode, through standard
conversion mechanisms, {\it e.g.} turn in the field trajectory, modulated reheating~\cite{Dvali:2003em,Kofman:2003nx} etc. We leave a detailed
exploration of these different conversion channels to future work.

\subsection{Quantum Corrections and Duration}

Moving beyond quadratic order, we next wish to assess the range of validity of the effective theory by looking at the various scales suppressing
higher-dimensional operators. The lowest such scale represents the most conservative value for the cutoff of the theory.  (By ``conservative", we mean the
cutoff which most severely limits the range of energy within which the effective field theory is valid.) It should be kept in mind, however, that some of these scales could be fake, for instance if there are field redefinitions that raise the cutoff~\cite{rub7}.  In this case, the true cutoff will be higher, so we can only err on the side of caution by reading the cutoff from the lagrangian.
This generalizes the flat-space analysis of Section~\ref{loops} to include mixing with gravity, and we will find that the resulting cutoff
is lowered considerably due to gravity. For this purpose, we focus on cubic interactions. 

Starting with the adiabatic sector, the cubic action for $\zeta$, up to a field redefinition and using  $a \simeq 1$, is given by~\cite{maldacena}
\bea
\nonumber
S_{3}  &\simeq & M_{\rm Pl}^2\int {\rm d}\tau {\rm d}^3x \Bigg\{\epsilon^2 \zeta \zeta'^2 + \epsilon^2 \zeta(\vec{\nabla}\zeta)^2
- 2\epsilon^2\zeta'\vec{\nabla} \zeta\cdot \frac{\vec{\nabla}}{\nabla^2} \zeta' + \frac{\epsilon}{2}\dot{\eta}\zeta^2 \zeta'   \\
&& \,\,\,\,\,\,\,\,\,\,\,\,\,\,\,\,\,\,\,\,\, + \frac{\epsilon^3}{2}\zeta' \vec{\nabla}\zeta\cdot\frac{\vec{\nabla}}{\nabla^2} \zeta' +\frac{\epsilon^3}{4}\nabla^2\zeta\left(\frac{\vec{\nabla}}{\nabla^2}\zeta'\right)^2
\Bigg\}\,,
\label{action4}
\eea
where spatial derivatives are contracted with the Euclidean metric $\delta_{ij}$, and $\eta$ as usual denotes the time derivative of the equation of state
\be
\eta \equiv \frac{1}{H} \frac{{\rm d}\ln \epsilon}{{\rm d}t} = 6\lambda t^2M_{\rm Pl}^2\,.
\label{eta}
\ee 
Translating to the canonically-normalized field $v \simeq M_{\rm Pl}\sqrt{2\epsilon} \zeta$, 
and treating the time-dependent coefficient in this field redefinition as approximately constant for sufficiently small space-time regions, the cubic action takes the form
\bea
S_{3}  &\simeq & \int {\rm d}\tau {\rm d}^3x \bigg\{\frac{\sqrt{\epsilon}}{2^{3/2}M_{\rm Pl}} v v'^2 + \ldots +  \frac{\dot{\eta}}{2^{5/2}M_{\rm Pl}\sqrt{\epsilon}} v^2v' + \frac{\epsilon^{3/2}}{2^{5/2}M_{\rm Pl}} v' \vec{\nabla}v \cdot\frac{\vec{\nabla}}{\nabla^2}v' + \ldots  \bigg\}\,,
\label{action5}
\eea
where the ellipses indicate terms suppressed by a similar scale as the operator preceding them.

Let us look at the different terms more closely. To begin with, the $\dot{\eta}$ term is renormalizable. It is a dimension-four operator with dimensionless coupling 
\be
\frac{\dot{\eta}}{2^{5/2}M_{\rm Pl}\sqrt{\epsilon}}  \sim \sqrt{\lambda}\ll 1\,,
\ee
where we have used~(\ref{epsilonE}) and~(\ref{eta}). Hence this operator is well within the perturbative regime.

The first and third terms in~(\ref{action5}), meanwhile, are non-renormalizable, suppressed respectively by $M_{\rm Pl}/\sqrt{\epsilon}$ and $M_{\rm Pl}/\epsilon^{3/2}$.
Since $\epsilon \gg 1$ during the phase of interest, the smaller of these two scales is the latter, which we therefore identify as the strong coupling scale of the adiabatic sector:
$\Lambda^{(1)} \equiv M_{\rm Pl}/\epsilon^{3/2}$. Using~(\ref{phiback2}) and~(\ref{epsilonE}) to write $\epsilon \simeq 9\lambda t^2M_{\rm Pl}^2 = 18 M_{\rm Pl}^2/\phi^2$, we obtain 
\be
\Lambda^{(1)} = \frac{M_{\rm Pl}}{\epsilon^{3/2}} \sim \frac{\phi^3}{M_{\rm Pl}^2} \,.
\ee
In particular, since $\phi < M_{\rm Pl}$ during the phase of interest, this is clearly smaller than the flat-space cutoff~(\ref{strongcoup}), as claimed.

As discussed in Section~\ref{loops}, consistency of the effective field theory requires that the characteristic frequency of our background solution, $\omega = \dot{\phi}/\phi = 1/(-t)$, be smaller
than the cutoff. Recall that in flat space, the analogous check~(\ref{flatspacecheck}) was automatically satisfied since $\lambda \ll 1$. But since
the cutoff is lowered with gravity, here we obtain a non-trivial constraint on the allowed range of $\phi$:
\be
\frac{\omega}{\Lambda^{(1)}} = \frac{M_{\rm Pl}^2}{\phi^3 (-t) } \sim \sqrt{\lambda}\frac{M_{\rm Pl}^2}{\phi^2} \,\lsim\, 1 \,.
\ee
In other words, the effective field theory is valid for
\be
\phi  \,\gsim\,  \phi_{\rm begin} \equiv \lambda^{1/4} M_{\rm Pl}\,.
\label{phibegin}
\ee
The phase of interest is therefore consistently described within the effective theory for $t \,\gsim\, t_{\rm begin}$, where
\be
t_{\rm begin} = \lambda^{-3/4} M_{\rm Pl}^{-1}\,.
\label{tbegin}
\ee
Combined with~(\ref{tend}), we see that our approximations are valid for a finite duration $t_{\rm begin}\,\lsim\, t \,\lsim\, t_{\rm end}$. 

Another way to understand~(\ref{phibegin}) physically is to consider the wavenumber $k$ of modes generated. 
Using the fact that $k = |t|^{-1} = \sqrt{\lambda}\phi/\sqrt{2}$ when a given $\chi$ mode freezes out, it is easily seen that demanding 
$k \,\lsim\,  \Lambda^{(1)}$ at freeze-out reproduces~(\ref{phibegin}). 

Next consider the $\chi$ sector, with action~(\ref{Schigrav}). For simplicity, let us set $\kappa = \xi = 0$ and work at quadratic order in $\chi$. 
In terms of the canonically normalized variables $v$ and $\hat\chi = \bar\phi\chi$, the cubic contributions from this sector are
\bea
\nonumber
S_\chi &\simeq& \frac{1}{2} \int {\rm d}^3 x {\rm d}\tau \Big[ \hat\chi'^2  - (\vec\nabla \hat\chi)^2 \Big] \\
&+& \int {\rm d}^3 x {\rm d}\tau\left\{ - \frac{v'}{2^{3/2}\sqrt{\epsilon} H M_{\rm Pl}}\hat\chi'^2 - \frac{\partial_i v}{\sqrt{2\epsilon}HM_{\rm Pl}} \hat\chi'\partial_i\hat\chi - \frac{\sqrt{\epsilon}}{M_{\rm Pl}}\frac{\partial_i}{\vec{\nabla}^2}v' \hat\chi'\partial_i\hat\chi  + \ldots  \right\}\,.
\label{schicubic}
\eea
As with~(\ref{action5}), this result holds for sufficiently small space-time regions, and the ellipses indicate terms suppressed by similar or higher scales.

The last term is suppressed by $\Lambda^{(1)}$. The first two terms, however, involve a new scale
\be
\Lambda^{(2)} = \left(\sqrt{2\epsilon}|H|M_{\rm Pl}\right)^{1/2} \sim \lambda^{1/4}\phi\,.
\ee
Whether this scale is lower or higher than $\Lambda^{(1)}$ clearly depends on the value of $\phi$. For $\phi \,\lsim\, \lambda^{1/8}M_{\rm Pl}$,
we have $\Lambda^{(1)} < \Lambda^{(2)}$, and the opposite is true for $\phi \,\gsim\, \lambda^{1/8}M_{\rm Pl}$.
Thus we are led to identify the cutoff of the theory as
\be
\Lambda \equiv \left\{\begin{array}{cl}
\phi^3/M_{\rm Pl}^2  \hspace{20pt}&\text{for}\hspace{10pt} \phi \,\lsim\, \lambda^{1/8}M_{\rm Pl} \,, \\ \\
\lambda^{1/4}\phi  \hspace{20pt}&\text{for}\hspace{10pt} \phi \,\gsim\, \lambda^{1/8}M_{\rm Pl} \,.
\end{array}\right.
\label{LambdaE}
\ee
(Note that since $\phi_{\rm begin} = \lambda^{1/4}M_{\rm Pl} \ll \lambda^{1/8}M_{\rm Pl}$, the arguments that led to $\phi_{\rm begin}$
in~(\ref{phibegin}) using $\Lambda^{(1)}$ as the cutoff still hold.) As mentioned earlier, it is possible there are field redefinitions or changes of gauge which can eliminate this scale, in which case the true cutoff is higher.  The value here is most conservative.

Having identified the cutoffs, we can do further checks of the regime of validity of the effective theory. 
The perturbation lagrangians~(\ref{action5}) and~(\ref{schicubic}) were derived under the approximation that $H$
and other background quantities are approximately constant. Following the same logic as the discussion above~(\ref{flatspacecheck}),
consistency of the effective theory requires that the Hubble parameter,
\be
H = \frac{1}{3\lambda t^3M_{\rm Pl}^2} \sim \sqrt{\lambda} \frac{\phi^3}{M_{\rm Pl}^2}\,,
\ee
be smaller than the cutoff. This is indeed the case:
\be
\frac{H}{\Lambda} \sim \left\{\begin{array}{cl}
\sqrt{\lambda} \ll 1  \hspace{20pt}&\text{for}\hspace{10pt} \phi \,\lsim\, \lambda^{1/8}M_{\rm Pl} \,, \\ \\
\lambda^{1/4}\phi^2/M_{\rm Pl}^2 \ll 1  \hspace{20pt}&\text{for}\hspace{10pt}M_{\rm Pl}\,\gsim\, \phi \,\gsim\, \lambda^{1/8}M_{\rm Pl} \,.
\end{array}\right.
\label{HEvscutoff}
\ee
Similarly, time derivatives of the Hubble parameter are also consistently small,
\be
\frac{1}{\Lambda^{n+1}} \frac{{\rm d}^n H}{{\rm d}t^n} \sim \frac{1}{(\Lambda t)^n} \frac{H}{\Lambda} \ll 1\,,
\ee
using the fact that $(\Lambda t)^{-1} \sim \sqrt{\lambda}M_{\rm Pl}^2/\phi^2  \ll 1$
for $\lambda^{1/4}M_{\rm Pl}\, \lsim \, \phi \,\lsim\, \lambda^{1/8}M_{\rm Pl}$ and $\sim \lambda^{1/4}\ll 1$ for $\phi \, \gsim\, \lambda^{1/8}M_{\rm Pl}$.

While $H$ and its time derivatives are smaller than the appropriate cutoffs, other background quantities need not be. For instance, the potential energy $\bar{V} \sim \lambda\bar\phi^4$
is clearly $\gg \Lambda^4$ at early times. In other words, although $\dot{H} \ll \Lambda^2$, the pressure is large
$P_\phi \sim M_{\rm Pl}^2\dot{H} \gg \Lambda^4$. This sounds worrisome, but the background quantities for the fields, about which the effective field theory is defined, need not themselves have small values below the cutoff.  To illustrate this point, consider the textbook case of the $SO(N)$ linear sigma model ${\cal L}=-{1\over 2}(\partial\Phi^I)^2+{1\over 2}\mu^2\Phi^{I2}-\frac{\lambda}{4}(\Phi^{I2})^2$,
for a multiplet of $N$ fields $\phi^I$.  Around the symmetry breaking vacuum $|\Phi|=\mu/\sqrt{\lambda}$ there is an effective field theory for the $N-1$ massless goldstone fields with cutoff $\Lambda\sim \mu$ obtained by integrating out the massive radial excitation of mass $\sqrt{2} \mu$.  On the other hand, the energy density of the vacuum, which is physical if we imagine weakly coupling to gravity, is $\mu^4/\lambda\gg \Lambda^4$.

\section{Observational Considerations}
\label{obscons}

In this Section, we discuss various phenomenological properties of the density perturbation spectrum derived above.
First of all, because our various approximations are valid for a finite time interval, $t_{\rm begin} \,\lsim\, t \,\lsim\, t_{\rm end}$,
with the end points given by~(\ref{tend}) and~(\ref{tbegin}), the range of scale invariant modes is also finite:
\be
\frac{k_{\rm end}}{k_{\rm begin}} = \frac{t_{\rm begin}}{t_{\rm end}} = \lambda^{-1/4}\,.
\label{rangelam}
\ee
Hence exponentially small values of $\lambda$ are required to obtain a sufficiently broad range of modes. 
Note that this entire range of modes is on super-Hubble scales by the end of the evolution:
\be
\frac{k}{|H(t_{\rm end})|} < \frac{k_{\rm end}}{|H(t_{\rm end})|} = 3\lambda t_{\rm end}^2M_{\rm Pl}^2 = 3 \,.
\label{kendsuper}
\ee
That is, the shortest-wavelength mode thus generated is comparable to the Hubble radius at the end of the evolution.
Hence it is reasonable to expect that the scale invariance of the spectrum will be preserved throughout the subsequent evolution,
independent of the details of the bounce and transition to the expanding, radiation-dominated phase.

At $t=t_{\rm end}$ our effective field theory description in terms of a negative quartic potential breaks down,
and we envision that higher-dimensional operators ${\cal O}(\phi^6/M_{\rm Pl}^2)$ modify the shape of the potential,
as sketched in Fig.~\ref{phi4pot}. While there is considerable freedom in specifying the subsequent evolution,
eventually the universe must bounce from contraction to expansion and reheat to a hot big bang phase.

Depending on the details of the transition, the Hubble parameter at the end of the scale invariant phase, $H(t_{\rm end})$,
will get related to $H_{\rm reheat}$, the Hubble parameter at the onset of the (expanding) radiation-dominated phase. Given such a relation, we can fix
$\lambda$ by demanding that the modes generated overlap with the largest observable scales today.
As a concrete example, suppose that $H_{\rm reheat}$ is comparable to $|H(t_{\rm end})|$:
\be
H_{\rm reheat} \simeq |H(t_{\rm end})| =  \frac{\sqrt{\lambda}M_{\rm Pl}}{3}\,,
\label{Hratiofix}
\ee
in which case the reheating temperature is
\be
T_{\rm reheat} \sim \sqrt{H_{\rm reheat}M_{\rm Pl}} \simeq \sqrt{|H(t_{\rm end})|M_{\rm Pl}}\sim \lambda^{1/4}M_{\rm Pl}\,.
\label{Trh}
\ee
Let $k_0$ and $k_{\rm reheat}$ denote the comoving wavenumber of the Hubble radius today
and at reheating, respectively. Assuming for simplicity that the universe is radiation-dominated all the way to the present epoch, then
\be
\frac{k_0}{k_{\rm reheat}} =  \frac{a_0H_0}{a_{\rm reheat}H_{\rm reheat}} \sim 10^{-30} \sqrt{\frac{M_{\rm Pl}}{|H(t_{\rm end})|}} \sim 10^{-30}\lambda^{-1/4}\,.
\label{k0krh}
\ee
In order for the scale invariant modes to overlap with the largest scales, we must therefore demand that
\be
\frac{k_{\rm end}}{k_{\rm begin}} \simeq \frac{k_0}{k_{\rm reheat}}\,. 
\ee
This requirement, combined with~(\ref{rangelam}) and~(\ref{k0krh}), fixes $\lambda$:
\be
\lambda \simeq 10^{-60}\,.
\label{lamfix}
\ee
Thus the theory is extremely weakly coupled. While this is certainly welcome from the point of view of perturbation theory,
an immediate worry is whether such small values of $\lambda$ are robust against radiative corrections.  We will show in Section~\ref{loopestimates} that loop corrections are indeed under control,
at least in the non-gravitational sector. 

Substituting~(\ref{lamfix}) into~(\ref{Trh}) fixes the reheating temperature to be at the weak scale:
\be
T_{\rm reheat}  \sim \lambda^{1/4}M_{\rm Pl} \sim {\rm TeV}\,. 
\ee
We stress that these specific results for $\lambda$ and $T_{\rm reheat}$ follow because of the assumption~(\ref{Hratiofix}).
Relaxing this condition would result in a range of allowed values for $\lambda$ and reheating temperature. 

Before closing this discussion, it is worth noting that the amplitude of $\chi$ perturbations is related to $H_{\rm end}$ as
\be
k^{3/2}|\chi_k| = \frac{3}{2} \frac{ |H(t_{\rm end})|}{M_{\rm Pl}}\,,
\ee
as can be seen by comparing~(\ref{chiampfinal}) and~(\ref{Hratiofix}). Thus the $\chi$ amplitude is set by the Hubble parameter, much like
a massless scalar field in de Sitter. Given our constraint that $\lambda \sim 10^{-60}$, this amplitude is of order $10^{-30}$, comparable to the amplitude
of a spectator scalar in TeV-scale inflation. To be observationally viable, the scenario requires a conversion mechanism that transfers this $10^{-30}$
entropic perturbation to the observed $10^{-5}$ adiabatic amplitude, analogous to a curvaton mechanism operating in TeV-scale inflation. 

It is also worth noting that the blue spectrum of $\zeta$ given by~(\ref{zetaampfinal}) is exponentially smaller than the $\chi$ amplitude on large scales.
Indeed, using the fact from~(\ref{kendsuper}) that the shortest-wavelength mode generated corresponds to $k_{\rm end} \sim |H_{\rm end}|$, 
we conclude from~(\ref{zetaampfinal}) that the $\zeta$ amplitude is comparable to the $\chi$ {\it on that scale}. But since the spectrum for $\zeta$ is very blue,
whereas that of $\chi$ is nearly scale invariant, the $\zeta$ amplitude is indeed negligible relative to $\chi$ on observable scales.

\section{Loop Corrections}
\label{loopestimates}

In this Section we estimate loop corrections to the classical action, focusing on the non-gravitational sector of the theory. Specifically, 
we do not consider mixing with gravity and work at the level of the flat-space perturbed action given by~(\ref{Lpert}), setting $\kappa=\xi = 0$ for simplicity:
\be
{\cal L} = -\frac{1}{2}(\partial\varphi)^2 + \frac{3}{2}\lambda \bar{\phi}^2\varphi^2 + \lambda\bar{\phi}\varphi^3 + \frac{\lambda}{4}\varphi^4  -\frac{1}{2}(\partial\hat{\chi})^2 
- \frac{\varphi}{\bar{\phi}} (\partial\hat{\chi})^2 - \frac{\varphi^2}{2\bar{\phi}^2}(\partial\hat{\chi})^2\,,
\label{Lpert2}
\ee
where $\hat\chi = \bar\phi\chi$ is the canonically-normalized field.

\begin{figure} 
   \centering
   \includegraphics[width=4.0in]{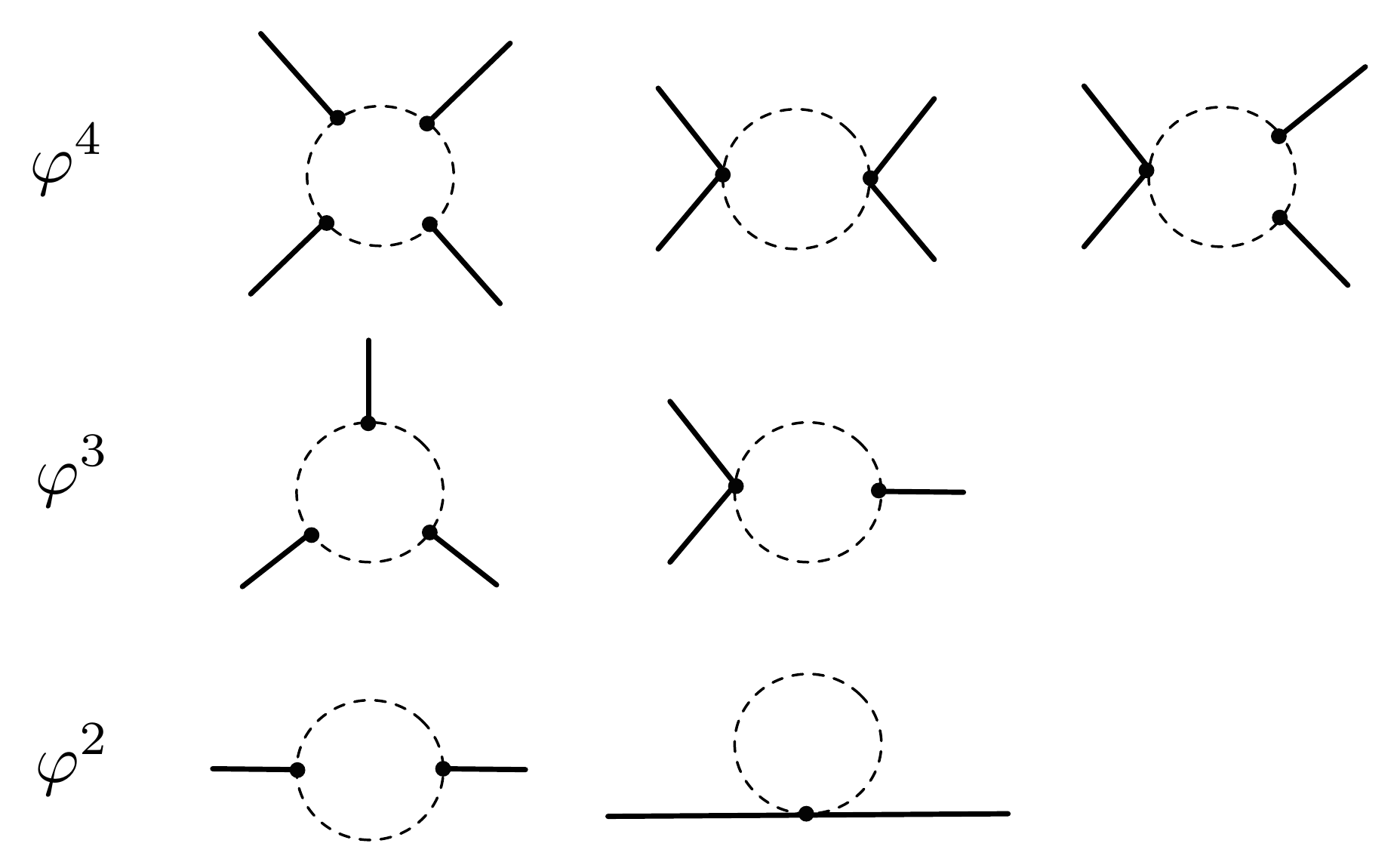}
   \caption{\textit{Loop contributions to the $\varphi$ sector, with $\hat\chi$ running in the loops. Solid lines are $\varphi$ and dotted lines are $\hat\chi$.  These graphs are potentially the most dangerous, since they do not involve $\lambda$ vertices.}}
      \label{loopcontributions}
\end{figure}

Let us start with the $\varphi$ sector. Since $\lambda$ must be exponentially small, as seen in Section~\ref{obscons}, it is clear that the most dangerous quantum corrections to 
this sector come from loop diagrams {\it without} any $\lambda$ vertices. That is, the dangerous diagrams involve $\hat\chi$ running in the loops. Fiducial radiative corrections of this type to the mass, cubic and quartic vertices are shown in Fig.~\ref{loopcontributions}. Power counting gives the following estimates:
\bea
\nonumber
&{\rm mass}:& \qquad \delta( \lambda \bar{\phi}^2 ) \sim   \frac{\Lambda^4}{\bar{\phi}^2} \,\lsim\, \lambda \bar{\phi}^2\,,  \\
\nonumber
&{\rm cubic}:& \qquad  \delta( \lambda \bar{\phi} ) \sim  \frac{\Lambda^4}{\bar{\phi}^3} \,\lsim\, \lambda \bar{\phi}\,, \\
&{\rm quartic}:& \qquad \delta\lambda \sim \frac{\Lambda^4}{\bar{\phi}^4} \,\lsim\, \lambda\,,
\eea
where we have used~(\ref{LambdaE}) for $\Lambda$. Hence loop corrections to $\varphi$ vertices are under control, thanks to the cutoff $\Lambda$ being suppressed by $\lambda$.

\begin{figure} 
   \centering
   \includegraphics[width=4.0in]{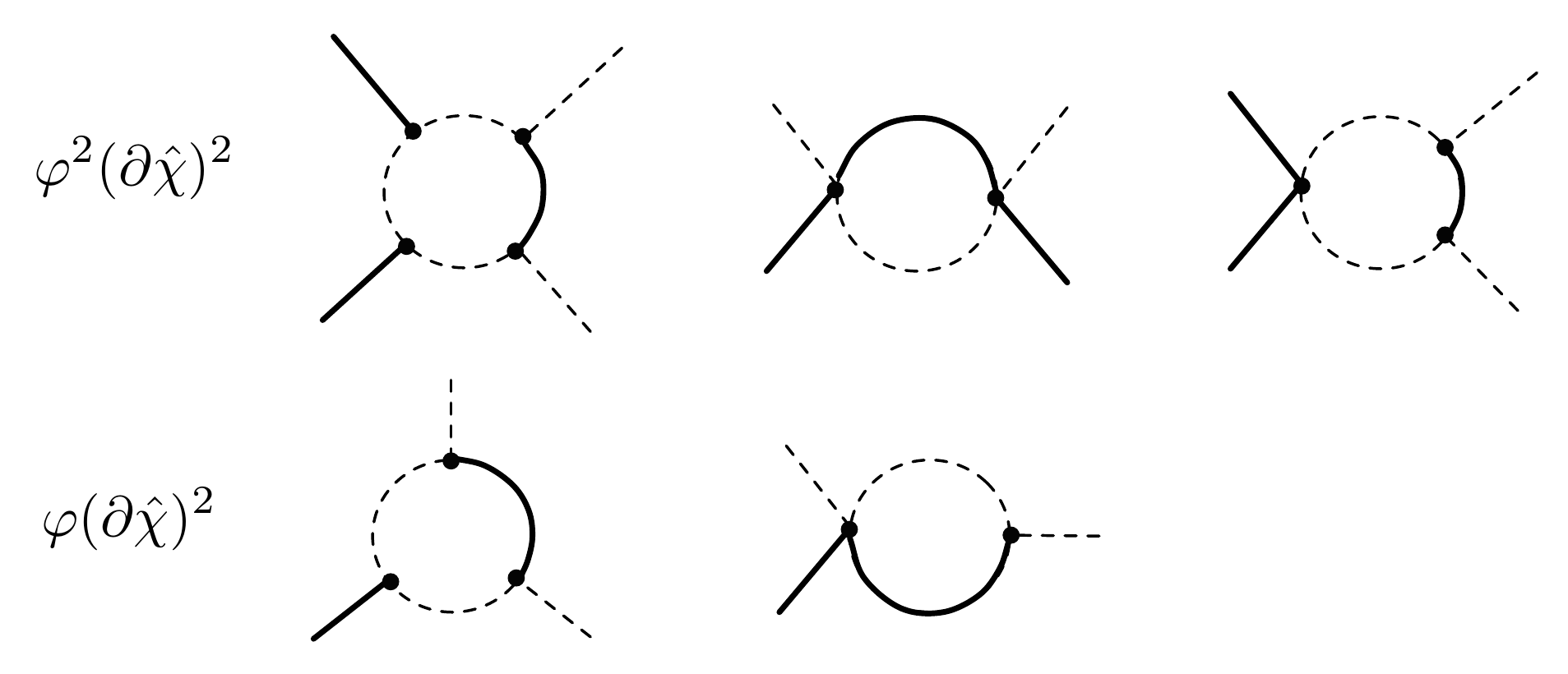}
   \caption{\textit{Dominant loop corrections to $\varphi-\hat\chi$ operators.  Solid lines are $\varphi$ and dotted lines are $\hat\chi$.} }
      \label{chiloops}
\end{figure}

The fiducial radiative corrections to the $\varphi(\partial\hat\chi)^2$ and  $\varphi^2(\partial\hat\chi)^2$ vertices are shown in Fig.~\ref{chiloops}.
It is easily seen that these corrections are suppressed by a power of $\sqrt{\lambda}$, and hence are completely negligible. For instance, the two diagrams
contributing to $\varphi(\partial\hat\chi)^2$ give
\be
\delta\left(\frac{\varphi}{\bar\phi}(\partial\hat\chi)^2\right) \sim \frac{\Lambda^2}{\bar\phi^2} \frac{\varphi}{\bar\phi}(\partial\hat\chi)^2 \,\lsim\, \sqrt{\lambda} \, \frac{\varphi}{\bar\phi}(\partial\hat\chi)^2 \ll  \frac{\varphi}{\bar\phi}(\partial\hat\chi)^2  \,.
\ee

One potentially dangerous contribution is a radiatively-generated mass term for $\hat\chi$, but loop corrections to the mass (and to any potential terms containing $\hat\chi$)  vanish due to the fact that $\hat\chi$ appears only with derivatives in~(\ref{Lpert2}) and so enjoys a shift symmetry.   We conclude that scale invariance
will be preserved under quantum contributions.   For the more general case of non-zero $\kappa$ and/or $\xi$, loops would generate a mass proportional to these parameters, hence small values of $\kappa$ and $\xi$ are technically natural.

\section{Other examples}
\label{confgal}

As emphasized many times already, our mechanism applies very generally --- {\it any} conformal field theory
that allows the fields to develop a $\bar\phi \sim 1/t^{d_I}$ background will naturally generate scale invariant perturbations for the weight 0 fields.
In this Section, we review some existing incarnations of the mechanism, which were already mentioned briefly in the Introduction.

The proposed symmetry breaking pattern arises in conformally invariant galileon theory~\cite{Nicolis:2008in}.
The scalar field lagrangian consists of a sum of derivative interaction terms,
\be
{\cal L}_{\rm gal} = c_2{\cal L}_2 +  c_3{\cal L}_3 +  c_4{\cal L}_4 +  c_5{\cal L}_5\,,
\label{Lconfgal}
\ee
where the coefficients $c_i$'s are dimensionless constants, and where\footnote{The galileon scalar field is denoted by $\pi$ in~\cite{Nicolis:2008in} and is related to our $\phi$ by $\phi \sim e^\pi$.}
\bea
{\cal L}_2 &=& -\frac{1}{2} (\pt \phi)^2\,, \nonumber \\
{\cal L}_3 &=& -\frac{1}{2}\frac{(\pt\phi)^2\Box\phi}{\phi^3} + \frac{1}{4}\frac{(\pt\phi)^4}{\phi^4}\,, \nonumber \\
{\cal L}_4 &=& \frac{(\pt\phi)^2}{\phi^4} \bigg[-\frac{1}{2}\frac{(\Box\phi)^2}{\phi^2} +\frac{1}{2}\frac{\phi^{,\mu\nu}\phi_{,\mu\nu}}{\phi^2} + \frac{6}{5}\frac{(\pt\phi)^2\Box\phi}{\phi^3} -\frac{6}{5}\frac{\phi^{,\mu}\phi^{,\nu}\phi_{,\mu\nu}}{\phi^3} -\frac{3}{20}\frac{(\pt\phi)^4}{\phi^4}\bigg]\,,  \nonumber \\
{\cal L}_5 &=& \frac{(\pt\phi)^2}{\phi^6} \bigg[-\frac{1}{2}\frac{(\Box\phi)^3}{\phi^3} -\frac{\phi^{,\mu\nu}\phi_{,\nu\rho}{\phi^{,\rho}}_\mu}{\phi^3} +\frac{3}{2}\frac{\Box\phi\phi^{,\mu\nu}\phi_{,\mu\nu}}{\phi^3} 
+ 3\frac{\phi^{,\mu\nu}\phi_{,\nu\rho}\phi^{,\rho}\phi_{,\mu}}{\phi^4} - 3\frac{\Box\phi \phi^{,\mu\nu}\phi_{,\mu}\phi_{,\nu}}{\phi^4} \nonumber\\
& - & 3\frac{(\partial\phi)^2\phi^{,\mu\nu}\phi_{,\mu\nu}}{\phi^4} + 3\frac{(\Box\phi)^2(\partial\phi)^2}{\phi^4} -\frac{36}{7}\frac{(\pt\phi)^4\Box\phi}{\phi^5} + \frac{36}{7}\frac{(\pt\phi)^2\phi^{,\mu}\phi^{,\nu}\phi_{,\mu\nu}}{\phi^5}  -\frac{3}{56}\frac{(\pt\phi)^6}{\phi^6} \bigg ]\,.
\label{confgalterms}
\eea
Each ${\cal L}_i$ individually transforms to a total derivative under~(\ref{delphiconfmany}) with $d_I=1$, hence the theory is conformally invariant for
arbitrary choice of the $c_i$'s.   Furthermore, the derivative structure is such that the equations of motion are second-order, despite the presence of higher order derivatives in the lagrangian.  This guarantees that no extra degrees of freedom will propagate around any background.  This theory can also be thought of as the non-relativistic limit of a brane probing a background AdS$_5$ geometry, where the brane and bulk contain all possible Lovelock invariants~\cite{deRham:2010eu}.

This theory allows for a solution $\bar\phi=\alpha/(-t)$, provided the coefficients satisfy
\be
\alpha c_2 - \frac{3}{2\alpha}c_3 + \frac{3}{2\alpha^3}c_4 - \frac{3}{4\alpha^5}c_5 = 0\,.
\label{backcond}
\ee
Moreover, requiring that the field not be a ghost around this background requires that~\cite{Nicolis:2008in}
\be
c_2 - \frac{3}{\alpha^2} c_3 + \frac{9}{2\alpha^4} c_4 - \frac{3}{\alpha^6} c_5 > 0\,.
\label{galstability}
\ee
These coefficients can furthermore be chosen so that the stress energy tensor for $\phi$ violates the NEC~\cite{Nicolis:2009qm}.
This possibility has been exploited in the Galileon Genesis scenario~\cite{galgen} to generate an expanding universe from an 
asymptotically flat initial state. Instead, the cosmological scenario described in Section~\ref{gravon} for the negative quartic potential clearly satisfies the NEC, since the scalar field
in this case has canonical kinetic term. As mentioned already, the resulting cosmology consists of a slowly contracting universe. A similar scenario
can be realized with the conformal galileon, by choosing the $c_i$'s appropriately such that the NEC is satisfied. We leave a detailed analysis to future work.

An unavoidable drawback of the conformal galileon, however, is that fluctuations propagate superluminally on backgrounds
that represent small deformations from the $1/t$ solution~\cite{galgen}. This applies independently of whether the NEC is violated or not, and
may pose a challenge for UV-completing such theories in local quantum field theories or perturbative string theories~\cite{nimaUVIR}. 
 
Our fiducial example is that of a canonical scalar field rolling down a negative quartic potential (Section~\ref{phi4}).
As mentioned earlier, this is closely related to the $U(1)$-invariant model proposed recently by Rubakov~\cite{rub1}$-$\cite{rub7}, but there are also some
differences. Rubakov considers a complex scalar field $\psi$ with $U(1)$-invariant negative quartic potential 
\be
{\cal L}_\psi = -\frac{1}{2}\partial_\mu\psi\partial^\mu\psi^* + \frac{\lambda}{4}|\psi|^4\,.
\label{Spsi}
\ee
In terms of polar fields variables, $\psi = \phi e^{i\chi}$, this becomes
\be
{\cal L}_\psi =  -\frac{1}{2}(\partial\phi)^2 + \frac{\lambda}{4}\phi^4 - \frac{1}{2} \phi^2(\partial\chi)^2\,,
\label{Spolar}
\ee
which matches~(\ref{Sphi}) and~(\ref{Lchi}) in the special case $\kappa=\xi = 0$.
It was emphasized in~\cite{rub1} that conformal invariance {\it and} $U(1)$ invariance are
both necessary to generate scale invariant perturbations in the axionic direction. However, the general discussion of
Section~\ref{symmbreak} in fact shows that this is part of a more general construction.  Scale invariance of the perturbations follows from conformal
invariance --- {\it any} scaling dimension 0 field with sufficiently small dimensionless couplings will acquire a Harrison-Zeldovich spectrum.

As mentioned briefly in the Introduction, there are important differences between the broader cosmological scenario presented in~\cite{rub1}
and the model proposed here.

\begin{itemize}

\item The scenario of~\cite{rub1} assumes that $\psi$ is coupled conformally to gravity. Instead, in Section~\ref{gravon} we coupled our scalar field $\phi$
minimally to gravity, thereby explicitly breaking conformal invariance at the $1/M_{\rm Pl}$ level.

\item The scalar $\psi$ is a spectator field in~\cite{rub1}, whose backreaction on the cosmological background is assumed negligible.
An essential component of our scenario is that $\phi$ drives the cosmological background.  As we have seen,
the resulting background is a slowly contracting universe, with very stiff equation of state $w \gg 1$. This phase is a cosmological attractor,
which makes the universe increasingly flat, homogeneous and isotropic.

\item Since $\psi$ is a spectator field in~\cite{rub1}, its fluctuations $\varphi_k$ represent (gauge-invariant) entropy perturbations. In particular,
the strongly-red tilted spectrum~(\ref{delphigrows}) is physical and potentially problematic on large scales~\cite{rub3}.
In our case, the fluctuations $\varphi_k$ represent adiabatic perturbations. We have seen that they in fact result in a very {\it blue} spectrum for the
curvature perturbation $\zeta$, which is therefore exponentially suppressed on large scales.

\end{itemize}

\section{\label{DBIsec}Non-Linear Extensions: The DBI Example}

Thus far we have demanded that the conformal field theory we start with linearly realize the conformal symmetry $so(4,2)$, and that it be broken only by a time dependent expectation value.  However, we now show that it is possible to realize our general scenario in a theory which non-linearly realizes the $so(4,2)$ symmetry to begin with.  In particular, the $so(4,2)$ need not necessarily even be a space-time conformal symmetry, but could be some other enlargement of Poincar\'e symmetry. This opens up a vast new regime of possibilities.

For example, consider the action of a 3-brane moving in an AdS$_5$ geometry. The 3-brane is described by world-volume scalar fields describing the embedding of the brane into the bulk.  Four of these can be eliminated by gauge fixing the world-volume reparametrization invariance, leaving one scalar to describe the transverse motion of the brane.  The action inherits the $so(4,2)$ isometries of the bulk AdS$_5$ which it probes, and these symmetries act non-linearly on the scalar.

The lowest order action in an effective field theory expansion is the DBI lagragian,
\be
{\cal L}_{\rm DBI} = -\phi^4\sqrt{1+ \frac{(\partial\phi)^2}{\phi^4}} + \left(1+ \frac{\lambda}{4}\right)\phi^4\,.
\label{SDBI}
\ee
The special case $\lambda = 0$ corresponds to a supersymmetric brane, thus by allowing for $\lambda > 0$ we depart
from the supersymmetric case. The $so(4,2)$ isometries act on the brane non-linearly. In addition to the linear Poincar\'e symmetries,
we have the following transformations:
\bea
\nonumber
\delta_D \phi &=& - (1 + x^\mu\partial_\mu)\phi \,, \\
\delta_{K_\mu} \phi &=& \left( -2x_\mu - 2x_\mu x^\nu\partial_\nu + x^2\partial_\mu - \frac{1}{2\phi^2}\partial_\mu\right)\phi\,.
\label{delphiDBI}
\eea
Evidently, the special conformal transformations act non-linearly in this case.

Note that in the ``non-relativistic" limit, $|\partial\phi| \ll \phi^2$, this action reduces
to~(\ref{Sphi}), and it is a remarkable fact that the commutation relations (and hence the symmetry group) are not altered in this limit.  In this sense, the DBI example represents a relativistic extension of our earlier framework.

Nevertheless, this theory also admits a $\phi \sim 1/t$ solution for any $\lambda > 0$.  Around this solution, the spatial special conformal generators are now broken, in contrast to our other examples.  Indeed, the equation of motion for a purely time-dependent profile, $\phi = \bar\phi(t)$, is given by
\be
- \frac{{\rm d}}{{\rm d}t}\left( \frac{\dot{\bar\phi}}{\sqrt{1- \dot{\bar\phi}^2/\bar\phi^4}}\right) - \frac{2\dot{\bar\phi}^2}{\bar\phi \sqrt{1- \dot{\bar\phi}^2/\bar\phi^4}}
+  4\bar\phi^3\left(1 + \frac{\lambda}{4} - \sqrt{1- \dot{\bar\phi}^2/\bar\phi^4} \right)  = 0\,.
\ee
It is easily seen that this equation admits a solution 
\be
\bar\phi(t) = \frac{1 + \lambda/4}{\sqrt{1 + \lambda/8}} \frac{\sqrt{2}}{\sqrt{\lambda}(-t)} \,.
\ee
As a consistency check, we recover~(\ref{phiback}) in the limit $\lambda \ll 1$. In this limit, the potential is nearly flat, hence the brane is moving slowly.
In the opposite limit $\lambda \rightarrow \infty$, $\bar\phi(t) \rightarrow 1/(-t)$ becomes independent of $\lambda$, corresponding to
ultra-relativistic brane motion. By studying perturbations it is straightforward to show that the growing mode solution is $\varphi \sim 1/t^2$, which, as in the canonical case,
represents a constant time shift of the background. We leave a detailed analysis of this relativistic example to future work~\cite{godfrey}.

\section{\label{conclusions}Conclusions}

The pseudo-conformal universe is a novel scenario of early universe cosmology based on approximate conformal symmetry.
Unlike inflation, gravity is unimportant, and the universe is approximately flat Minkowski space for most of the evolution. The fields in the 
conformal field theory acquire a time-dependent background which breaks the $so(4,2)$ conformal algebra down to the $so(4,1)$
algebra of de Sitter space. The salient features of the scenario are completely determined by this symmetry breaking pattern, irrespective
of the underlying microphysics. In particular, conformal weight 0 fields acquire a scale invariant spectrum under very general conditions.
Including gravity, the scale factor is very slowly contracting, and the universe becomes increasingly flat, homogeneous and isotropic, thereby addressing
the traditional problems of standard big bang cosmology without invoking a phase of exponential expansion. 

The scenario makes distinguishing predictions from inflation. Since space-time is nearly static, gravitational waves are not appreciably excited, 
as in ekpyrotic cosmology~\cite{Khoury:2001wf,lathamgw}. Tensor modes preserve their adiabatic vacuum normalization, $h_k \sim 1/\sqrt{k}$,
which result in a strongly blue-tilted tensor spectrum
\be
P_h(k) \sim  k^3|h_k|^2 \sim k^2\,,
\ee
corresponding to $n_{\rm T} = 2$. The primordial tensor amplitude is therefore exponentially suppressed on observable scales. The dominant gravitational wave background at long wavelengths is the secondary gravitational waves induced by the energy density fluctuations, roughly $10^{-5}$ times smaller than the primordial fluctuations~\cite{second}.  
Detection of primordial gravitational waves, for instance through cosmic microwave background B-mode polarization~\cite{CMBPOL}, would unequivocally rule out this mechanism.

Since our mechanism for generating scale invariant perturbations employs multiple scalar fields, one expects significant non-gaussianities of the local
shape, potentially detectable by the Planck satellite and future experiments~\cite{Komatsu:2009kd}. The amplitude of the 3-point function depends on the details of the conversion mechanism from entropy perturbations to the adiabatic mode. To make concrete predictions, in upcoming work we will survey the landscape of possibilities by exploring different conversion channels,
such as turns in the field space trajectory and modulated reheating~\cite{Dvali:2003em,Kofman:2003nx}. 

A critical ingredient for the completeness of our scenario is a bounce from contraction to expansion. 
In ongoing work, we are investigating how to include a non-singular bouncing phase that derives from
consistent 4d effective field theories, as proxies for more ``realistic" bounces in string theory.
There are two known classes of field theories that can violate the NEC without ghosts
or other pathologies: the ghost condensate~\cite{ArkaniHamed:2003uy,Creminelli:2006xe} and the already-mentioned conformal galileon theories~\cite{Nicolis:2009qm,Creminelli:2006xe}. 
(See~\cite{Khoury:2010gb,Khoury:2011da} for a recent construction of the ${\cal N} =1$ supersymmetric extensions of these
theories.) Whether these theories can be UV completed has been questioned~\cite{nimaUVIR,dubovsky},
but from a cosmological perspective they offer consistent, low-energy effective field theories to study stable bouncing scenarios. 
This strategy was followed successfully in the New Ekpyrotic scenario~\cite{Buchbinder:2007ad,Creminelli:2007aq, Buchbinder:2007tw,Buchbinder:2007at}, which merged an ekpyrotic contracting phase to 
ghost-condensate phase bridging smoothly to an expanding phase. See~\cite{Lin:2010pf} for a similar approach in the case of the matter-bounce scenario~\cite{dust,Cai:2011zx}. The long-wavelength scale invariant spectrum generated during the
contracting phase was found to go through the bounce unscathed.

There is one potential advantage that our scenario has over inflation, which is that it may possess a 5d dual through AdS/CFT.  The early universe in our scenario is not described by an inflating de Sitter space (for which is has proven difficult to find any kind of dual theory) but rather a conformal field theory in approximately flat space, and AdS/CFT is precisely suited for studying strongly coupled CFTs in flat space.  The dual will be a gravitational theory on AdS$_5$, and the duality should be precise to the extent that gravitational $1/M_{\rm Pl}$ corrections can be ignored in the CFT.  This is the case for most of the evolution up until the end.  

The $so(4,2)$ conformal invariance of the CFT is dual to the $so(4,2)$ isometries of the AdS$_5$ theory.  The $\sim 1/ t^{d_I}$ expectation value which breaks the $so(4,2)$ symmetry of the CFT down to $so(4,1)$ should correspond in the bulk to some configuration which has the effect of foliating the AdS$_5$ into 4d de Sitter slices.  This foliation will then break the bulk isometry group down to the $so(4,1)$ isometries of the dS$_4$ slices.

Our scenario may be generalized to include gauge fields as well as scalars, and then AdS/CFT opens up many new possibilities for explicit realizations.  In this paper, we focused on the example of a massless scalar $\phi^4$ theory, which is conformally invariant at the classical level but possesses a quantum conformal anomaly.  The same is true of any classically conformal theory whose beta function does not vanish, such as QCD with massless quarks.  Through AdS/CFT constructions, however, there is no shortage of examples of gauge theories which are exactly conformally invariant at the quantum level.  If solutions with the right symmetry breaking pattern could be found in these examples, it would provide an ideal setting for our scenario, one which could even be studied at strong coupling from the AdS side.  A guiding example is the double-trace deformation of ${\cal N} = 4$ super Yang-Mills studied by~\cite{Turok:2007ry,Craps:2007ch}.

\textit{Acknowledgments:} It is our pleasure to thank Vijay Balasubramanian, Ben Craps, Paolo Creminelli, Austin Joyce, Godfrey Miller, Alberto Nicolis, Burt Ovrut, Zain Saleem, Leonardo Senatore, Paul Steinhardt, James Stokes, Andrew Tolley and Mark Trodden for helpful discussions. We are especially grateful to Valery Rubakov for offering detailed and insightful feedback on an earlier version of the manuscript. This work was supported in part by funds from the University of Pennsylvania, the Alfred P. Sloan Foundation and NSF grant PHY-0930521.


\begin{thebibliography}{99}

\bibitem{inf} 
  A.~A.~Starobinsky,
  ``Relict Gravitation Radiation Spectrum and Initial State of the Universe.
  (In Russian),''
  JETP Lett.\  {\bf 30}, 682 (1979);
 A.~H.~Guth,
  ``The Inflationary Universe: A Possible Solution To The Horizon And Flatness
  Problems,''
  Phys.\ Rev.\  D {\bf 23}, 347 (1981);
    K.~Sato,
  ``First Order Phase Transition of a Vacuum and Expansion of the Universe,''
  Mon.\ Not.\ Roy.\ Astron.\ Soc.\  {\bf 195}, 467 (1981);
  A.~Albrecht and P.~J.~Steinhardt,
  ``Cosmology For Grand Unified Theories With Radiatively Induced Symmetry
  Breaking,''
  Phys.\ Rev.\ Lett.\  {\bf 48}, 1220 (1982);
  A.~D.~Linde,
  ``A New Inflationary Universe Scenario: A Possible Solution Of The Horizon,
  Flatness, Homogeneity, Isotropy And Primordial Monopole Problems,''
  Phys.\ Lett.\  B {\bf 108}, 389 (1982).

\bibitem{Brandenberger:1988aj}
  R.~H.~Brandenberger and C.~Vafa,
  ``Superstrings in the Early Universe,''
  Nucl.\ Phys.\  B {\bf 316}, 391 (1989).

\bibitem{Nayeri:2005ck}
  A.~Nayeri, R.~H.~Brandenberger and C.~Vafa,
  ``Producing a scale-invariant spectrum of perturbations in a Hagedorn  phase
  of string cosmology,''
  Phys.\ Rev.\ Lett.\  {\bf 97}, 021302 (2006)
  [arXiv:hep-th/0511140].

\bibitem{Brandenberger:2006xi}
  R.~H.~Brandenberger, A.~Nayeri, S.~P.~Patil and C.~Vafa,
  ``Tensor modes from a primordial Hagedorn phase of string cosmology,''
  Phys.\ Rev.\ Lett.\  {\bf 98}, 231302 (2007)
  [arXiv:hep-th/0604126].
  
\bibitem{Brandenberger:2006vv}
  R.~H.~Brandenberger, A.~Nayeri, S.~P.~Patil and C.~Vafa,
  ``String gas cosmology and structure formation,''
  Int.\ J.\ Mod.\ Phys.\  A {\bf 22}, 3621 (2007)
  [arXiv:hep-th/0608121].

\bibitem{Brandenberger:2006pr}
  R.~H.~Brandenberger {\it et al.},
  ``More on the Spectrum of Perturbations in String Gas Cosmology,''
  JCAP {\bf 0611}, 009 (2006)
  [arXiv:hep-th/0608186].
  
\bibitem{Battefeld:2005av}
  T.~Battefeld and S.~Watson,
  ``String gas cosmology,''
  Rev.\ Mod.\ Phys.\  {\bf 78}, 435 (2006)
  [arXiv:hep-th/0510022].

\bibitem{Brandenberger:2008nx}
  R.~H.~Brandenberger,
  ``String Gas Cosmology,''
  arXiv:0808.0746 [hep-th].

\bibitem{Gasperini:1992em}
  M.~Gasperini and G.~Veneziano,
  ``Pre - big bang in string cosmology,''
  Astropart.\ Phys.\  {\bf 1}, 317 (1993)
  [arXiv:hep-th/9211021].

\bibitem{Gasperini:2002bn}
  M.~Gasperini and G.~Veneziano,
  ``The Pre - big bang scenario in string cosmology,''
  Phys.\ Rept.\  {\bf 373}, 1 (2003)
  [arXiv:hep-th/0207130].

\bibitem{Gasperini:2007vw}
  M.~Gasperini and G.~Veneziano,
  ``String Theory and Pre-big bang Cosmology,''
  arXiv:hep-th/0703055.


\bibitem{Khoury:2001wf}
  J.~Khoury, B.~A.~Ovrut, P.~J.~Steinhardt and N.~Turok,
  ``The ekpyrotic universe: Colliding branes and the origin of the hot big
  bang,''
  Phys.\ Rev.\  D {\bf 64}, 123522 (2001)
  [arXiv:hep-th/0103239].

\bibitem{Donagi:2001fs}
  R.~Y.~Donagi, J.~Khoury, B.~A.~Ovrut, P.~J.~Steinhardt and N.~Turok,
  ``Visible branes with negative tension in heterotic M-theory,''
  JHEP {\bf 0111} (2001) 041
  [arXiv:hep-th/0105199].

\bibitem{Khoury:2001bz}
  J.~Khoury, B.~A.~Ovrut, N.~Seiberg, P.~J.~Steinhardt and N.~Turok,
  ``From big crunch to big bang,''
  Phys.\ Rev.\  D {\bf 65}, 086007 (2002)
  [arXiv:hep-th/0108187].

\bibitem{Khoury:2001zk}
  J.~Khoury, B.~A.~Ovrut, P.~J.~Steinhardt and N.~Turok,
  ``Density perturbations in the ekpyrotic scenario,''
  Phys.\ Rev.\  D {\bf 66}, 046005 (2002)
  [arXiv:hep-th/0109050].

\bibitem{Lyth:2001pf}
  D.~H.~Lyth,
  ``The primordial curvature perturbation in the ekpyrotic universe,''
  Phys.\ Lett.\  B {\bf 524}, 1 (2002)
  [arXiv:hep-ph/0106153].

\bibitem{Brandenberger:2001bs}
  R.~Brandenberger and F.~Finelli,
  ``On the spectrum of fluctuations in an effective field theory of the
  ekpyrotic universe,''
  JHEP {\bf 0111}, 056 (2001)
  [arXiv:hep-th/0109004].

\bibitem{Steinhardt:2001st}
  P.~J.~Steinhardt and N.~Turok,
  ``Cosmic evolution in a cyclic universe,''
  Phys.\ Rev.\  D {\bf 65}, 126003 (2002)
  [arXiv:hep-th/0111098].

\bibitem{Notari:2002yc}
  A.~Notari and A.~Riotto,
  ``Isocurvature perturbations in the ekpyrotic universe,''
  Nucl.\ Phys.\  B {\bf 644}, 371 (2002)
  [arXiv:hep-th/0205019].

\bibitem{Finelli:2002we}
  F.~Finelli,
  ``Assisted contraction,''
  Phys.\ Lett.\  B {\bf 545}, 1 (2002)
  [arXiv:hep-th/0206112].

\bibitem{Tsujikawa:2002qc}
  S.~Tsujikawa, R.~Brandenberger and F.~Finelli,
  ``On the construction of nonsingular pre-big-bang and ekpyrotic cosmologies
  and the resulting density perturbations,''
  Phys.\ Rev.\  D {\bf 66}, 083513 (2002)
  [arXiv:hep-th/0207228].
 
\bibitem{Gratton:2003pe}
  S.~Gratton, J.~Khoury, P.~J.~Steinhardt and N.~Turok,
  ``Conditions for generating scale-invariant density perturbations,''
  Phys.\ Rev.\  D {\bf 69}, 103505 (2004)
  [arXiv:astro-ph/0301395].

\bibitem{Tolley:2003nx}
  A.~J.~Tolley, N.~Turok and P.~J.~Steinhardt,
  ``Cosmological perturbations in a big crunch / big bang space-time,''
  Phys.\ Rev.\  D {\bf 69}, 106005 (2004)
  [arXiv:hep-th/0306109].
  
\bibitem{Craps:2003ai}
  B.~Craps and B.~A.~Ovrut,
  ``Global fluctuation spectra in big crunch / big bang string vacua,''
  Phys.\ Rev.\  D {\bf 69}, 066001 (2004)
  [arXiv:hep-th/0308057].

\bibitem{Khoury:2003vb}
  J.~Khoury, P.~J.~Steinhardt and N.~Turok,
  ``Great expectations: Inflation versus cyclic predictions for spectral
  tilt,''
  Phys.\ Rev.\ Lett.\  {\bf 91}, 161301 (2003)
  [arXiv:astro-ph/0302012].

\bibitem{Khoury:2003rt}
  J.~Khoury, P.~J.~Steinhardt and N.~Turok,
  ``Designing Cyclic Universe Models,''
  Phys.\ Rev.\ Lett.\  {\bf 92}, 031302 (2004)
  [arXiv:hep-th/0307132].

\bibitem{Khoury:2004xi}
  J.~Khoury,
  ``A briefing on the ekpyrotic / cyclic universe,''
  arXiv:astro-ph/0401579.

\bibitem{Creminelli:2004jg}
  P.~Creminelli, A.~Nicolis and M.~Zaldarriaga,
  ``Perturbations in bouncing cosmologies: Dynamical attractor vs scale
  invariance,''
  Phys.\ Rev.\  D {\bf 71}, 063505 (2005)
  [arXiv:hep-th/0411270].

\bibitem{Lehners:2007ac}
  J.~L.~Lehners, P.~McFadden, N.~Turok and P.~J.~Steinhardt,
  ``Generating ekpyrotic curvature perturbations before the big bang,''
  Phys.\ Rev.\  D {\bf 76}, 103501 (2007)
  [arXiv:hep-th/0702153].

\bibitem{Buchbinder:2007ad}
  E.~I.~Buchbinder, J.~Khoury and B.~A.~Ovrut,
  ``New Ekpyrotic Cosmology,''
  Phys.\ Rev.\  D {\bf 76}, 123503 (2007)
  [arXiv:hep-th/0702154].


\bibitem{Creminelli:2007aq}
  P.~Creminelli and L.~Senatore,
  ``A smooth bouncing cosmology with scale invariant spectrum,''
  JCAP {\bf 0711}, 010 (2007)
  [arXiv:hep-th/0702165].

\bibitem{Buchbinder:2007tw}
  E.~I.~Buchbinder, J.~Khoury and B.~A.~Ovrut,
  ``On the Initial Conditions in New Ekpyrotic Cosmology,''
  JHEP {\bf 0711}, 076 (2007)
  [arXiv:0706.3903 [hep-th]].

\bibitem{Buchbinder:2007at}
  E.~I.~Buchbinder, J.~Khoury and B.~A.~Ovrut,
  ``Non-Gaussianities in New Ekpyrotic Cosmology,''
  Phys.\ Rev.\ Lett.\  {\bf 100}, 171302 (2008)
  [arXiv:0710.5172 [hep-th]].

\bibitem{Koyama:2007mg}
  K.~Koyama and D.~Wands,
  ``Ekpyrotic collapse with multiple fields,''
  JCAP {\bf 0704}, 008 (2007)
  [arXiv:hep-th/0703040].

\bibitem{Koyama:2007ag}
  K.~Koyama, S.~Mizuno and D.~Wands,
  ``Curvature perturbations from ekpyrotic collapse with multiple fields,''
  Class.\ Quant.\ Grav.\  {\bf 24}, 3919 (2007)
  [arXiv:0704.1152 [hep-th]].

\bibitem{Lehners:2007wc}
  J.~L.~Lehners and P.~J.~Steinhardt,
  ``Non-Gaussian Density Fluctuations from Entropically Generated Curvature
  Perturbations in Ekpyrotic Models,''
  Phys.\ Rev.\  D {\bf 77}, 063533 (2008)
  [Erratum-ibid.\  D {\bf 79}, 129903 (2009)]
  [arXiv:0712.3779 [hep-th]].

\bibitem{Lehners:2008my}
  J.~L.~Lehners and P.~J.~Steinhardt,
  ``Intuitive understanding of non-Gaussianity in ekpyrotic and cyclic
  models,''
  Phys.\ Rev.\  D {\bf 78}, 023506 (2008)
  [Erratum-ibid.\  D {\bf 79}, 129902 (2009)]
  [arXiv:0804.1293 [hep-th]].

\bibitem{Lehners:2009qu}
  J.~L.~Lehners and P.~J.~Steinhardt,
  ``Non-Gaussianity Generated by the Entropic Mechanism in Bouncing Cosmologies
  Made Simple,''
  Phys.\ Rev.\  D {\bf 80}, 103520 (2009)
  [arXiv:0909.2558 [hep-th]].

\bibitem{Khoury:2009my}
  J.~Khoury and P.~J.~Steinhardt,
  ``Adiabatic Ekpyrosis: Scale-Invariant Curvature Perturbations from a Single
  Scalar Field in a Contracting Universe,''
  Phys.\ Rev.\ Lett.\  {\bf 104}, 091301 (2010)
  [arXiv:0910.2230 [hep-th]].

\bibitem{Linde:2009mc}
  A.~Linde, V.~Mukhanov and A.~Vikman,
  ``On adiabatic perturbations in the ekpyrotic scenario,''
  JCAP {\bf 1002}, 006 (2010)
  [arXiv:0912.0944 [hep-th]].
  
\bibitem{Khoury:2010gw}
  J.~Khoury, G.~E.~J.~Miller,
  ``Towards a Cosmological Dual to Inflation,''
    [arXiv:1012.0846 [hep-th]].
  
\bibitem{Khoury:2011ii}
  J.~Khoury and P.~J.~Steinhardt,
  ``Generating Scale-Invariant Perturbations from Rapidly-Evolving Equation of
  State,''
  arXiv:1101.3548 [hep-th], accepted for publication in Phys.\ Rev.\  D.

\bibitem{Joyce:2011ta}
  A.~Joyce and J.~Khoury,
  ``Scale Invariance via a Phase of Slow Expansion,''
  arXiv:1104.4347 [hep-th].

\bibitem{Lehners:2008vx}
  J.~L.~Lehners,
  ``Ekpyrotic and Cyclic Cosmology,''
  Phys.\ Rept.\  {\bf 465}, 223 (2008)
  [arXiv:0806.1245 [astro-ph]].
  
\bibitem{Lehners:2010fy}
  J.~L.~Lehners,
  ``Ekpyrotic Non-Gaussianity -- A Review,''
  Adv.\ Astron.\  {\bf 2010}, 903907 (2010)
  [arXiv:1001.3125 [hep-th]].

\bibitem{Wands:1998yp}
  D.~Wands,
  ``Duality invariance of cosmological perturbation spectra,''
  Phys.\ Rev.\  D {\bf 60}, 023507 (1999)
  [arXiv:gr-qc/9809062].

\bibitem{dust}
  F.~Finelli and R.~Brandenberger,
  ``On the generation of a scale invariant spectrum of adiabatic fluctuations
  in cosmological models with a contracting phase,''
  Phys.\ Rev.\  D {\bf 65}, 103522 (2002)
  [arXiv:hep-th/0112249].

\bibitem{Brandenberger:2010dk}
  R.~H.~Brandenberger,
  ``Cosmology of the Very Early Universe,''
  AIP Conf.\ Proc.\  {\bf 1268}, 3 (2010)
  [arXiv:1003.1745 [hep-th]].

\bibitem{Cai:2011zx}
  Y.~F.~Cai, R.~Brandenberger and X.~Zhang,
  ``The Matter Bounce Curvaton Scenario,''
  JCAP {\bf 1103}, 003 (2011)
  [arXiv:1101.0822 [hep-th]].

\bibitem{AdSCFT}
  O.~Aharony, S.~S.~Gubser, J.~M.~Maldacena, H.~Ooguri and Y.~Oz,
  ``Large N field theories, string theory and gravity,''
  Phys.\ Rept.\  {\bf 323}, 183 (2000)
  [arXiv:hep-th/9905111].

\bibitem{rub1}
  V.~A.~Rubakov,
  ``Harrison-Zeldovich spectrum from conformal invariance,''
  JCAP {\bf 0909}, 030 (2009)
  [arXiv:0906.3693 [hep-th]].

\bibitem{rub2}
  M.~Osipov and V.~Rubakov,
  ``Scalar tilt from broken conformal invariance,''
  JETP Lett.\  {\bf 93}, 52 (2011)
  [arXiv:1007.3417 [hep-th]].

\bibitem{rub3}
M.~Libanov and V.~Rubakov,
  ``Cosmological density perturbations from conformal scalar field: infrared
  properties and statistical anisotropy,''
  JCAP {\bf 1011}, 045 (2010)
  [arXiv:1007.4949 [hep-th]].
  
\bibitem{rub4}
  M.~Libanov, S.~Mironov and V.~Rubakov,
  ``Properties of scalar perturbations generated by conformal scalar field,''
  arXiv:1012.5737 [hep-th].
  
\bibitem{rub5}
  M.~Libanov, S.~Ramazanov and V.~Rubakov,
  ``Scalar perturbations in conformal rolling scenario with intermediate
  stage,''
  arXiv:1102.1390 [hep-th].

\bibitem{rub6}
  M.~Libanov, S.~Mironov and V.~Rubakov,
  ``Non-Gaussianity of scalar perturbations generated by conformal
  mechanisms,''
  arXiv:1105.6230 [astro-ph.CO].

\bibitem{rub7}
  M.~Libanov, V.~Rubakov,
  ``Dynamical vs spectator models of (pseudo-)conformal Universe,''
  [arXiv:1107.1036 [hep-th]].


\bibitem{galgen}
  P.~Creminelli, A.~Nicolis and E.~Trincherini,
  ``Galilean Genesis: An Alternative to inflation,''
  JCAP {\bf 1011}, 021 (2010)
  [arXiv:1007.0027 [hep-th]].

\bibitem{nimaUVIR}
  A.~Adams, N.~Arkani-Hamed, S.~Dubovsky, A.~Nicolis and R.~Rattazzi,
  ``Causality, analyticity and an IR obstruction to UV completion,''
  JHEP {\bf 0610}, 014 (2006)
  [arXiv:hep-th/0602178].




\bibitem{Erickson:2003zm}
  J.~K.~Erickson, D.~H.~Wesley, P.~J.~Steinhardt and N.~Turok,
  ``Kasner and mixmaster behavior in universes with equation of state $w \geq 1$,''
  Phys.\ Rev.\  D {\bf 69}, 063514 (2004)
  [arXiv:hep-th/0312009].

\bibitem{Garfinkle:2008ei}
  D.~Garfinkle, W.~C.~Lim, F.~Pretorius and P.~J.~Steinhardt,
  ``Evolution to a smooth universe in an ekpyrotic contracting phase with $w > 1$,''
  Phys.\ Rev.\  D {\bf 78}, 083537 (2008)
  [arXiv:0808.0542 [hep-th]].

\bibitem{mottola1}
  I.~Antoniadis, P.~O.~Mazur and E.~Mottola,
  ``Conformal invariance and cosmic background radiation,''
  Phys.\ Rev.\ Lett.\  {\bf 79}, 14 (1997)
  [arXiv:astro-ph/9611208].

\bibitem{mottola2}
  I.~Antoniadis, P.~O.~Mazur and E.~Mottola,
  ``Conformal Invariance, Dark Energy, and CMB Non-Gaussianity,''
  arXiv:1103.4164 [gr-qc].

\bibitem{Khoury:2008wj}
  J.~Khoury and F.~Piazza,
  ``Rapidly-Varying Speed of Sound, Scale Invariance and Non-Gaussian
  Signatures,''
  JCAP {\bf 0907}, 026 (2009)
  [arXiv:0811.3633 [hep-th]].

\bibitem{Baumann:2011dt}
  D.~Baumann, L.~Senatore and M.~Zaldarriaga,
  ``Scale-Invariance and the Strong Coupling Problem,''
  JCAP {\bf 1105}, 004 (2011)
  [arXiv:1101.3320 [hep-th]].

\bibitem{csm}
  J.~Magueijo and J.~Noller, 
  ``Primordial fluctuations without scalar fields,"
  Phys.\ Rev.\  D {\bf 81}, 043509 (2010).
  
\bibitem{csk}
  D.~Bessada, W.~H.~Kinney, D.~Stojkovic and J.~Wang,
  ``Tachyacoustic Cosmology: An Alternative to Inflation,"
  Phys.\ Rev.\  D {\bf 81}, 043510 (2010).
  
\bibitem{csfedo}
  J.~Magueijo, J.~Noller and F.~Piazza,
  ``Bimetric structure formation: non-Gaussian predictions,"
   Phys.\ Rev.\  D {\bf 82}, 043521 (2010). 

\bibitem{csearlier}
   C.~Armendariz-Picon and E.~A.~Lim,
   ``Scale Invariance without Inflation?,"
   JCAP {\bf 0312}, 002 (2003),
     C.~Armendariz-Picon,
  ``Near Scale Invariance with Modified Dispersion Relations,"
  JCAP {\bf 0610}, 010 (2006).

\bibitem{Piao:2006ja}
  Y.~-S.~Piao,
  ``Seeding Primordial Perturbations During a Decelerated Expansion,''
  Phys.\ Rev.\  {\bf D75}, 063517 (2007).
  [gr-qc/0609071].

\bibitem{Piao:2008ip}
  Y.~-S.~Piao,
  ``On Primordial Density Perturbation and Decaying Speed of Sound,''
  Phys.\ Rev.\  {\bf D79}, 067301 (2009).
  [arXiv:0807.3226 [gr-qc]].


\bibitem{Tolley:2007nq}
  A.~J.~Tolley and D.~H.~Wesley,
  ``Scale-invariance in expanding and contracting universes from two-field
  models,''
  JCAP {\bf 0705}, 006 (2007)
  [arXiv:hep-th/0703101].

\bibitem{Low:2001bw}
  I.~Low and A.~V.~Manohar,
  ``Spontaneously broken space-time symmetries and Goldstone's theorem,''
  Phys.\ Rev.\ Lett.\  {\bf 88}, 101602 (2002)
  [arXiv:hep-th/0110285].

\bibitem{Coleman:1969sm}
  S.~R.~Coleman, J.~Wess and B.~Zumino,
  ``Structure of phenomenological Lagrangians. 1,''
  Phys.\ Rev.\  {\bf 177}, 2239 (1969).

\bibitem{Callan:1969sn}
  C.~G.~Callan, S.~R.~Coleman, J.~Wess and B.~Zumino,
  ``Structure of phenomenological Lagrangians. 2,''
  Phys.\ Rev.\  {\bf 177}, 2247 (1969).

\bibitem{Hinterbichler:2012mv} 
  K.~Hinterbichler, A.~Joyce and J.~Khoury,
  ``Non-linear Realizations of Conformal Symmetry and Effective Field Theory for the Pseudo-Conformal Universe,''
  arXiv:1202.6056 [hep-th].

\bibitem{volkov}
D.~V.~Volkov, 
``Phenomenological Lagrangians,''
Sov.\ J.\ Particles\ Nucl. {\bf 4}, 3 (1973).

\bibitem{Isham:1970xz}
  C.~J.~Isham, A.~Salam and J.~A.~Strathdee,
  ``Broken chiral and conformal symmetry in an effective-lagrangian
  formalism,''
  Phys.\ Rev.\  D {\bf 2}, 685 (1970).

\bibitem{Salam:1970qk}
  A.~Salam and J.~A.~Strathdee,
  ``Nonlinear realizations. 2. Conformal symmetry,''
  Phys.\ Rev.\  {\bf 184}, 1760 (1969).

\bibitem{Isham:1970gz}
  C.~J.~Isham, A.~Salam and J.~A.~Strathdee,
  ``Spontaneous breakdown of conformal symmetry,''
  Phys.\ Lett.\  B {\bf 31}, 300 (1970).

\bibitem{Coleman:1973jx}
  S.~R.~Coleman and E.~J.~Weinberg,
  ``Radiative Corrections as the Origin of Spontaneous Symmetry Breaking,''
  Phys.\ Rev.\  D {\bf 7}, 1888 (1973).

\bibitem{Weinberg:1996kr}
  S.~Weinberg,
  ``The quantum theory of fields. Vol. 2: Modern applications,''
{\it  Cambridge, UK: Univ. Pr. (1996) 489 p}

\bibitem{Weinberg:1987vp}
  E.~J.~Weinberg and A.~q.~Wu,
  ``Understanding Complex Perturbative Effective Potentials,''
  Phys.\ Rev.\  D {\bf 36}, 2474 (1987).

\bibitem{Nicolis:2008in}
  A.~Nicolis, R.~Rattazzi and E.~Trincherini,
  ``The Galileon as a local modification of gravity,''
  Phys.\ Rev.\  D {\bf 79}, 064036 (2009)
  [arXiv:0811.2197 [hep-th]].

\bibitem{Dvali:2003em}
  G.~Dvali, A.~Gruzinov and M.~Zaldarriaga,
  ``A new mechanism for generating density perturbations from inflation,''
  Phys.\ Rev.\  D {\bf 69}, 023505 (2004)
  [arXiv:astro-ph/0303591].

\bibitem{Kofman:2003nx}
  L.~Kofman,
  ``Probing string theory with modulated cosmological fluctuations,''
  arXiv:astro-ph/0303614.


\bibitem{maldacena}
  J.~M.~Maldacena,
  ``Non-Gaussian features of primordial fluctuations in single field
  inflationary models,''
  JHEP {\bf 0305}, 013 (2003)
  [arXiv:astro-ph/0210603].


\bibitem{deRham:2010eu}
  C.~de Rham and A.~J.~Tolley,
  ``DBI and the Galileon reunited,''
  JCAP {\bf 1005}, 015 (2010)
  [arXiv:1003.5917 [hep-th]].

\bibitem{Nicolis:2009qm}
  A.~Nicolis, R.~Rattazzi and E.~Trincherini,
  ``Energy's and amplitudes' positivity,''
  JHEP {\bf 1005}, 095 (2010)
  [arXiv:0912.4258 [hep-th]].

\bibitem{godfrey}
K.~Hinterbichler, J.~Khoury and G.~Miller, to appear.

\bibitem{lathamgw}
  L.~A.~Boyle, P.~J.~Steinhardt and N.~Turok,
  ``The cosmic gravitational wave background in a cyclic universe,''
  Phys.\ Rev.\  D {\bf 69}, 127302 (2004)
  [arXiv:hep-th/0307170].

\bibitem{second}
  D.~Baumann, P.~J.~Steinhardt, K.~Takahashi and K.~Ichiki,
  ``Gravitational Wave Spectrum Induced by Primordial Scalar Perturbations,''
  Phys.\ Rev.\  D {\bf 76}, 084019 (2007)
  [arXiv:hep-th/0703290].

\bibitem{CMBPOL}
  D.~Baumann {\it et al.}  [CMBPol Study Team Collaboration],
  ``CMBPol Mission Concept Study: Probing Inflation with CMB Polarization,''
  AIP Conf.\ Proc.\  {\bf 1141}, 10 (2009)
  [arXiv:0811.3919 [astro-ph]].

\bibitem{Komatsu:2009kd}
  E.~Komatsu {\it et al.},
  ``Non-Gaussianity as a Probe of the Physics of the Primordial Universe and
  the Astrophysics of the Low Redshift Universe,''
  arXiv:0902.4759 [astro-ph.CO].

\bibitem{ArkaniHamed:2003uy}
  N.~Arkani-Hamed, H.~C.~Cheng, M.~A.~Luty and S.~Mukohyama,
  ``Ghost condensation and a consistent infrared modification of gravity,''
  JHEP {\bf 0405}, 074 (2004)
  [arXiv:hep-th/0312099].

\bibitem{Creminelli:2006xe}
  P.~Creminelli, M.~A.~Luty, A.~Nicolis and L.~Senatore,
  ``Starting the universe: Stable violation of the null energy condition and
  non-standard cosmologies,''
  JHEP {\bf 0612}, 080 (2006)
  [arXiv:hep-th/0606090].

\bibitem{Khoury:2010gb}
  J.~Khoury, J.~L.~Lehners and B.~Ovrut,
  ``Supersymmetric P(X,phi) and the Ghost Condensate,''
  arXiv:1012.3748 [hep-th].
  
\bibitem{Khoury:2011da}
  J.~Khoury, J.~L.~Lehners and B.~A.~Ovrut,
  ``Supersymmetric Galileons,''
  arXiv:1103.0003 [hep-th].

\bibitem{dubovsky}
  S.~L.~Dubovsky and S.~M.~Sibiryakov,
  ``Spontaneous breaking of Lorentz invariance, black holes and perpetuum
  mobile of the 2nd kind,''
  Phys.\ Lett.\  B {\bf 638}, 509 (2006)
  [arXiv:hep-th/0603158].

\bibitem{Lin:2010pf}
  C.~Lin, R.~H.~Brandenberger and L.~P.~Levasseur,
  ``A Matter Bounce By Means of Ghost Condensation,''
  JCAP {\bf 1104}, 019 (2011)
  [arXiv:1007.2654 [hep-th]].

\bibitem{Turok:2007ry}
  N.~Turok, B.~Craps and T.~Hertog,
  ``From big crunch to big bang with AdS/CFT,''
  arXiv:0711.1824 [hep-th].

\bibitem{Craps:2007ch}
  B.~Craps, T.~Hertog and N.~Turok,
  ``Quantum Resolution of Cosmological Singularities using AdS/CFT,''
  arXiv:0712.4180 [hep-th].
  

\end{thebibliography}
\end{document}